\documentclass[journal]{IEEEtran}
\usepackage{tabularx}
\usepackage{cite}
\usepackage{amsmath,amssymb,amsfonts}
\usepackage{algorithmic}
\usepackage{graphicx}
\usepackage{textcomp}

\usepackage{cite}
\usepackage{amsmath,amssymb,amsfonts}
\usepackage{algorithmic}
\usepackage{graphicx}
\usepackage{textcomp}
\usepackage{xcolor}
\usepackage{grfext}
\usepackage{kvoptions}
\usepackage{kvsetkeys}
\usepackage{comment} 

\usepackage{graphicx}
\usepackage[table,xcdraw]{xcolor}
\usepackage{colortbl} 

\usepackage{collectbox}
\usepackage{ltxcmds}
\usepackage{cite}
\usepackage{subfigure}
\usepackage{amsmath,amssymb}
\usepackage{textcomp}
\usepackage[normalem]{ulem}
\usepackage{epstopdf}
\usepackage[export]{adjustbox}

\begin{document}
\title{Device Identification and Motion Classification of Passive Nonlinear RF Transponders Using Wi-Fi Channel State Information}

\author{Cory Hilton, \IEEEmembership{Student Member, IEEE}, Sheng Huang, \IEEEmembership{Member, IEEE} \\and Jeffrey A. Nanzer, \IEEEmembership{Senior Member, IEEE}}

\maketitle

\begin{abstract}
A method of joint device authentication and gesture classification using low-cost passive transponders with signals of opportunity from commodity hardware is presented for the tracking of objects in cluttered environments. The proposed approach detects nonlinear responses of diode-based radio-frequency (RF) transponders and estimates the mean time-varying phase across all subcarriers channel state information (CSI) obtained from standard 802.11 preambles on commodity hardware. The measurement of these unique responses is enabled by passive RF transponders designed using 2.4 GHz meander-line monopole antennas with capacitive top hat loading, along with a low-barrier Schottky diode acting as the port between the two elements, enabling low-power omnidirectional sensing and complex gesture recognition with any $N$ number of unique RF transponders. The system for capturing these responses consists of a wireless access-point (AP) as the signal signal source, illuminating the RF transponder with a standard preamble which is then modulated by the nonlinear response of the diode and transmitted back to a single ESP32 receiver where the channel response is calculated. Using the modulated CSI response, device identification via RF fingerprinting is performed with simple machine learning models, achieving $\geq$95\% accuracy in both static and dynamic conditions. In the case of a transponder moving dynamically, the time-varying velocity profile is calculated using a novel phase sanitation approach, significantly reducing the impacts of the non-cooperative network. The approach is shown to enable high-accuracy estimation of micro-motions on a single receiver with a theoretical limit of 9.75 $\text{m}/\text{s}$ using an amplitude invariant binary comparator form of the sanitized waveform. The concept is simulated and experimentally validated, where lightweight deep-learning models built on a convolutional neural network and long short term memory (CNN-LSTM) architecture are then applied to a diverse dataset of five transponders and three motion profiles performed with a linear actuator, demonstrating a classification accuracy of $\approx$98\% in both gesture classification and device identification tasks. 
\end{abstract}


\section{Introduction}
\label{sec: Introduction}

As the internet of thing (IoT) expands, the usable frequencies within the existing 802.11 bands has become increasingly congested, requiring a growing network of wirelessly connected devices to seek new methods of utilizing the available bandwidth. \cite{wen2026bodyofrevolution,EM_Localization} Among these growing uses is the rapidly growing topic on integrated sensing and communications (ISAC), seeking to re-purpose signals of opportunity (SoO) for existing and emerging multi-variate sensing while maintaining integrity of the communications channels between devices \cite{riet2024developing}. In particular, accurate device identification, whether or not smart functionality is integrated into the device itself, is crucial for applications such as human-computer interfacing (HCI) \cite{ma2019wifi,FallDetection,CSI_USER}, contraband detection \cite{appleby2007standoff,gaire2026staticsense}, and home health monitoring \cite{wang2012simultaneous,alqaness2019channel,li2013review} among others. The principle approach to smart sensors has traditionally relied on radio-frequency identification (RFID)\cite{RFID_Review,harmonicRFID_bikedetection} implemented at a dedicated frequency band using either active or passive chipless radio-frequency (RF) transponders to receive and retransmit signals, involving unique spectra or data responses stored on the local circuitry \cite{chipless_rfid_ML,dipole,RFID_Localization}. Typically, these transponders are active, encoding information about the target onto the on-transponder circuitry, enabling the transponder to store information such as seen in near-field communication (NFC) devices \cite{RFID_Positioning,RFID_Localization}. The information from these devices can then impart information about the monitored target to specialized readers, enabling high-accuracy device authentication \cite{ShengRFFingerprinting,CSI_USER}. However, the complexity of these transponders has led to increased interest in low-cost, passive devices for short-range smart home detection and classification.

\begin{figure}[t!]
    \centering
    \includegraphics[width=\linewidth]{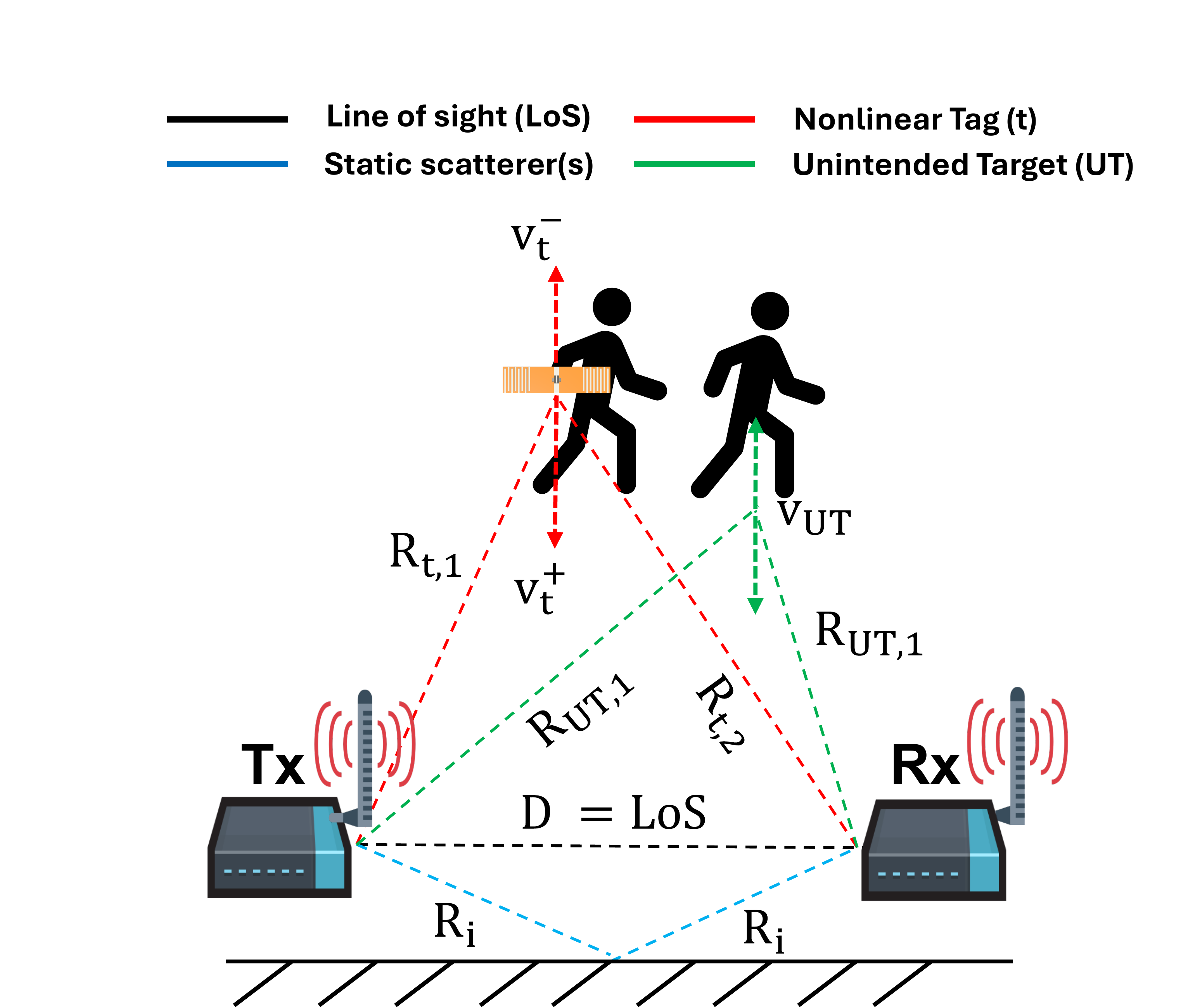}
    \caption{System model depicting the joint gesture recognition and RF fingerprinting topology. The wireless access point (Tx) illuminates the environment, and a single receiver (Rx) captures the nonlinear multipath reflections from the passive transponder and unintended targets moving at distinct velocities.}
    \label{fig: OV1}
\end{figure}

\begin{figure*}
    \centering
    \includegraphics[width=\linewidth]{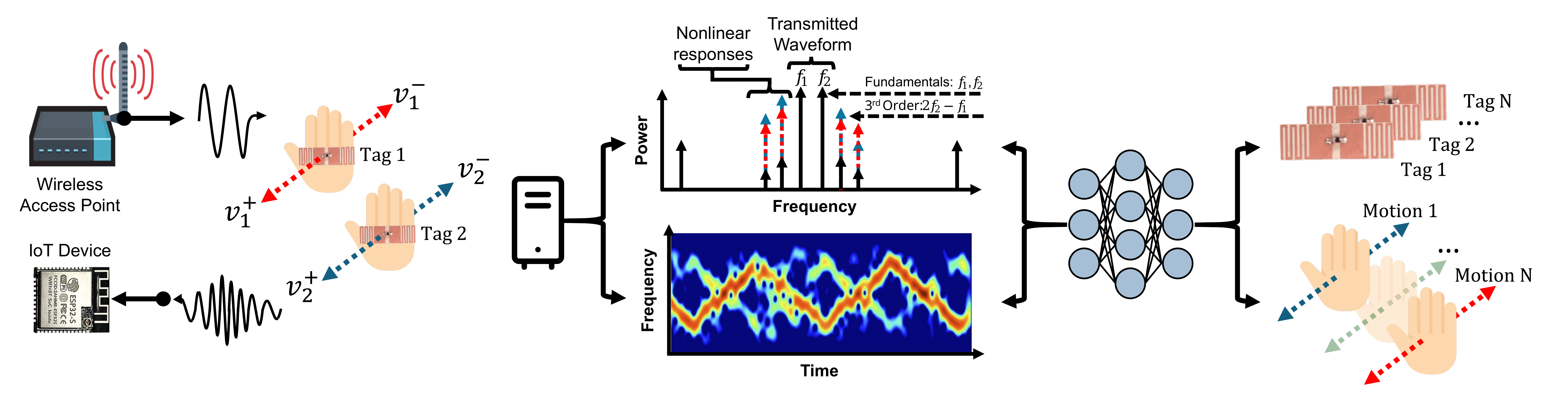}
    \caption{Proposed processing pipeline for joint device identification and gesture classification. Signals of opportunity are backscattered by the nonlinear tags, generating unique inter-modulation products (IMPs) and time-varying frequency shifts that are captured, processed into channel state information (CSI), and fed into a machine learning model.}
    \label{fig: OV1_Large}
\end{figure*}

In this work, we present a novel method for joint device identification and gesture classification using low-cost passive nonlinear transponders illuminated by known Wi-Fi preambles (see Figs. \ref{fig: OV1} and \ref{fig: OV1_Large}). By exploiting the nonlinear inter-modulation products (IMPs) generated by a Schottky diode integrated into the transponder, we demonstrate the ability to capture unique spectral fingerprints in addition to target-specific kinematic information~\cite{Strack,CSI_USER,ma2019wifi,alqaness2019channel}. Unlike traditional multi-node radar setups, our system leverages channel state information (CSI) from standard 802.11 preambles acquired via a single commercial off-the-shelf (COTS) ESP32 receiver \cite{ThruWallESP}. We introduce a reference-based phase noise reduction approach that enables accurate micro-Doppler velocity estimation on a single non-cooperative node, achieving theoretical limits of up to $9.75~\text{m}/\text{s}$. Furthermore, we demonstrate that lightweight machine learning models can achieve high classification accuracy in distinguishing both the unique devices and their respective dynamic motion profiles.

The remainder of this paper is organized as follows. Section~\ref{sec: Theory Diode} discusses the theoretical foundations of nonlinear backscattering and the generation of IMPs using multi-tone OFDM signals. Section~\ref{sec:csi velocity} details the methodology for extracting time-varying phase information from CSI, including the proposed phase correction and velocity extraction techniques. In Section~\ref{sec: System}, the design of the passive nonlinear transponder and the experimental CSI micro-Doppler radar system are presented. Section~\ref{sec: Measurement} outlines the experimental measurement procedures for both static RF fingerprinting and dynamic velocity estimation. Section~\ref{sec: Classification} evaluates the performance of various machine learning algorithms in jointly authenticating devices and classifying gestures. Finally, Section~\ref{sec: Conclusion} concludes the paper and discusses potential future work.

\section{Nonlinear Backscattering with Signals of Opportunity}
\label{sec: Theory Diode}

Nonlinear RF fingerprinting has been gaining interest as an alternative to active RFID due to the tags being lower cost, simpler, and passive without the need for energy or information storage. 
When implemented using simple nonlinear devices to leverage the device-specific nonlinear responses of practical components, the transponders can be made with a wide range of antennas and COTS parts.

\subsection{RF Fingerprinting Using Intermodulation Products}

In this section, we demonstrate the expected response of common Wi-Fi preambles such as the legacy-long training field (L-LTF) when operating in the presence of a passive nonlinear transponder (such as that described in~\cite{ShengRFFingerprinting}). Simple nonlinear transponders use two antennas connected by a nonlinear component such as a Schottky diode to introduce nonlinearities into the retransmitted signal. The inter-modulation products (IMPs) generated by the diode are dependent in minor but detectable ways on manufacturing difference from device to device, generating a unique spectral response on the amplitude and phase of signals passed through the diode, which has been shown to be useful for device identification~\cite{Hilton_CAMA,ShengRFFingerprinting}. The RF fingerprinting for these passive transponders is facilitated when the transponder receives the known transmitted waveform, modulates the response due to the Schottky diode that acts as the port between the two antennas, and retransmits the modulated waveform. 
The nonlinearity of the RF diode mainly includes the carrier transport nonlinearity and the junction capacitance nonlinearity. 
The voltage-current (\(V\text{-}I\)) characteristics of the PN diode follows the small signal model of the Shockley equation  
\begin{equation}
I = I_s \left( e^{\frac{V}{\eta V_T}} - 1 \right)
\label{eq:diode_equation}
\end{equation}
where \(I_{s}\) is the reverse-saturation current, $V_T$ is the thermal voltage, generally 26 mV, $V$ is the voltage across the diode, and $\eta$ is the ideality factor determining the transfer function and thus, the harmonic output \cite{sze2021physics}. For small signals, the current in the diode can be described using a Taylor series expansion to allow for numerical estimation of the harmonic components as 

\begin{equation}
I\approx I_{s}\sum _{m=1}^{\infty }\frac{1}{m!}\left(\frac{V}{\eta V_T}\right)^{m}\
\end{equation}

where the nonlinearity coefficient of each term is determined by the saturation current and ideality factor, and high-order terms are those that contribute directly to the nonlinearity characteristics. To apply this approach to classical communications signals, we model the signal at the input of the transponder to be some multi-tone OFDM signal, which can be defined on a set of $K$ subcarriers as
\begin{equation}
V = \sum _{k\in K}A_{k}\,e^{j2\pi k\Delta f(t-t_{g})}
\end{equation}
where $A_k$ are the predefined amplitudes assigned to each subcarrier index $k\Delta f = f_k$ at subcarrier offset $\Delta f$, and $t_{g}$ is the guard interval for each symbol. Substituting this signal into the expanded Schockley equation for $V_D$ then yields the expansion 
\begin{equation}
I\approx I_{s}\sum _{m=1}^{\infty }\frac{1}{m!}\left(\frac{1}{\eta V_T} \sum _{k\in K}A_{k}\,e^{j2\pi k\Delta f(t-t_{g})}  \right)^{m}\
\label{Shockley}
\end{equation}
The IMPs of any multi-tone waveform increase at harmonically related frequencies due to the nonliearity of the diode. Thus, manufacturing tolerances that slightly change the diode nonlinearites, retransmitted third-order IMP response will include minor nonlinear variations that are unique to each device. The responses at any receiver capable of observing the modulation may then be used to uniquely identify the transponders based on the differences of these nonlinearities from device to device.

\subsection{Channel State Information}

Ideally, the receiver would be capable of sampling the harmonic structure of the retransmitted waveform with a high sample rate over long periods of time. However, such capabilities are generally limited in practical, low-cost hardware that is deployed in IoT and smart home applications. Modern approaches using network interface cards (NICs) are equipped with capable modems, but lack the low-level functionality required to observe the raw spectral response, thus, another approach is required. 

For analysis of the multi-tone L-LTF waveform, information about the propagation of the signals captured in each frequency band is condensed into channel state information (CSI). CSI is commonly considered as a measure of the stability of a communications link that defines the signal properties at the receiver for a given carrier frequency or OFDM channel. CSI also provides a measure of the impacts of multi-path, including amplitude attenuation and phase shifts. However, the signals are discretely sampled responses that are limited by the modem of the receiver. A more general approach is to represent the impulse response as the convolution of the transmitted and received signals in the frequency domain, the channel frequency response CFR, which can be simply defined for any $M$ number of scatterers as~\cite{SmartWireless}
\begin{equation}
H(f) = \sum_{m=1}^{M}\alpha_m e^{-j2 \pi f \tau_m}
\end{equation}
where $\alpha$ and $\tau$ are the complex amplitude and phase delays of the $m^{th}$ path.
when discretely sampled as a function of the magnitude and phase error for the given CFR, CSI is obtained. These values are then reported by the receiving
Typical receivers measure the magnitude $|H|$ and the phase $\angle H$ of CSI per channel $k$, giving $H(f_k) = |H(f_k)|e^{\angle H(f_k)}$. 

CSI is generally non-deterministic, but for the purposes of evaluating the impact on motion estimation and classification, assuming simple environmental scenarios with a small number of strong scatterers is often a useful approach.
In complex environments with $M$ scatterers the responses at frequency channels $f_k$, will be proportional to both the cluttered environment scattering amplitudes and the harmonic amplitudes generated by a non-linear scatterer. The CSI can then be given by
\begin{equation}
H[k] = {\sum_{m=1}^{M}{\alpha_{k,m}  e^{-j2\pi f_{k,m} \tau_m} }}
\label{eq: CSI}
\end{equation}
where $f_{k,m}$ represents the IMPs relative to the frequency $f_k$ for the $m$th scatterer. 

\begin{figure}[t!]
    \centering
    \includegraphics[width=\linewidth]{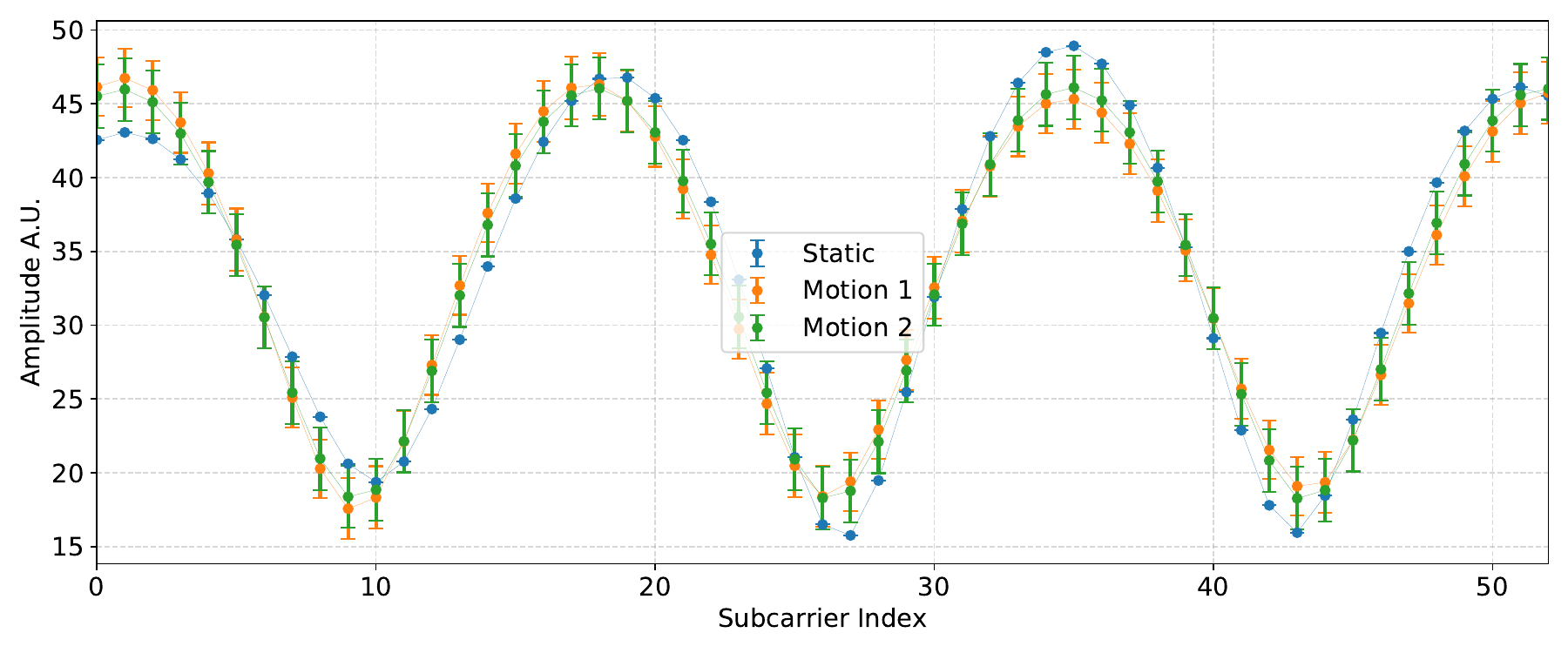}
    \caption{Simulated CSI amplitude across 52 subcarriers, demonstrating the frequency-selective fading and nulls where target motion exhibits the highest sensitivity to spatial displacements.}
    \label{fig: Sim CSI Fingerprint}
\end{figure}

Generally, simulation  of CSI is associated with a high computation cost due to the requirement to perform accurate path tracing of cluttered environments. However, in the simple cases with a small number of static time delays, the relative amplitudes and phase shifts of the small number of propagating signals can be estimated and more easily modeled. For a small number of strong scatterers, the interaction of the signals results in frequency selective fading, as shown in Fig.~\ref{fig: Sim CSI Fingerprint}. 
Nulls can be introduced in the spectrum based on the frequencies and delays of the propagating signals, and the CFR can be given as
\begin{equation}
H(f) = \alpha_1 + \alpha_2 e^{-j(2\pi f \Delta \tau_2 - \Delta \phi_2)}+...\alpha_M e^{-j(2\pi f \Delta \tau_M - \Delta \phi_M)}
\end{equation}
The frequency of the first nulled subcarrier can be calculated from the CSI phase term according to $f_{\text{null}}~=~\frac{1}{2\Delta\tau}(1+~\frac{\Delta\phi}{\pi})$.

\section{CSI Velocity Estimation}
\label{sec:csi velocity}

Micro-Doppler velocity estimation has been long studied as a method of measuring the motion of desired targets based on a time-varying radial range relative to a both bistatic and monostatic radar. 
In distributed wireless networks the transmitters and receivers are generally not colocated and thus bistatic operation should be considered.
The instantaneous frequency response is generally resolved via the time-derivative of received signal phase scaled by the target velocity vector and observation angles relative to boresight of the baseline, giving the bistatic Doppler shift (see e.g.,~\cite{chen2019micro})
\begin{equation}
f_{D,b} = \frac{2f}{c}\left|\textbf{v}\right| \text{cos}\left(\frac{\beta}{2}\right) \text{cos}(\delta)
\label{BiDoppler}
\end{equation}
where $\textbf{v}$ is the velocity vector of the target, $\beta$ is the angle between line-of-sight for the transmitter and receiver to the target, and $\delta$ is the the aspect angle between the target velocity vector and the relative observation angle from the centroid of the channel baseline. Therefore, for a target moving radially central to the baseline of the transmitter and receiver, the change in the Doppler frequency relative to a monostatic radar is $\Delta f_{D} = \text{cos}(\beta/2)$ in typical systems. Unfortuantely, low-cost wireless NICs such as the ESP32 generally do not provide access to direct RF sampling at the modem, limiting the available information for estimating velocity. However, because the CSI  varies as a function of frequency and channel, dynamic motion of objects in the field of view generate amplitude and phase modulation of the CSI as a function of time and frequency. This modulation can be used to estimate the motion of objects in the scene in a manner similar to typical micro-Doppler estimation. 
With a dynaically moving object in the field of view, the CSI can be considered the superposition of a series of static and dynamic scatterers, $H(t) = H_S(\tau^s)+H_D(\tau(t))$. The CSI can then be considered as a function of the time-varying range to the target, which is proportional to the instantaneous Doppler frequency $f_{D,b}(t)$, which can be extrapolated from the propagation delay $\tau$ as 
\begin{align*}
H(t,k)&=\sum_{m_\text{S}=1}^{M_\text{S}} \alpha^\text{S}_{k,m_\text{S}}(t)e^{-j2\pi f_{k,m_\text{S}}\tau^s_{k,m}- \Delta \phi_m)} \\
&+\sum_{m_\text{D}=1}^{M_\text{D}} \alpha^\text{D}_{k,m_\text{D}} e^{j2\pi f_{D,k,m_\text{D}} t}
\end{align*}
The raw CSI thus captures the complex baseband frequency response of the wireless channel modulated by the multipath fading of the nonlinear transponder and other scatterers. Let $H(t,k) \in \mathbb{C}^{N \times K}$ represent this raw CSI matrix, where $N$ denotes the number of captured time samples (packets) and $K$ denotes the number of available subcarriers.

\subsection{Phase Correction for Improved Doppler Estimation}

A common problem with commercial off-the-shelf (COTS) RF transceivers, especially when operating in non-cooperative networks, are the unsynchronized oscillators, leading to Carrier Frequency Offset (CFO) and Sampling Frequency Offset (SFO), both of which complicate frequency estimation, and thus velocity estimation.
The frequency offsets manifest as a random phase drift across each received CSI value, which masks the small phase variations relative to micro-motions that are induced by target movement. Where previous methods have utilized the CSI amplitude to determine the target velocity for single-receiver systems, we mitigate this phase error in a similar manner to existing multi-node approaches by introducing a reference-based phase noise mitigation approach. Referencing the instantaneous phase of a designated subcarrier as  $\phi_{\text{ref}} = \angle H[n_r,k]$, the phase drift across all subcarriers is assumed to be uniform such that phase gradient mitigation can be applied. Thus subtracting the reference phase mitigates the phase across all remaining subcarriers at every time instance $n$, $H[n,k] e^{-j \phi_{\text{ref}}[n]}$.
This results in each subcarrier only retaining phase information relative to the dynamic motion of the target at every sample $[n,k]$, or rather the Doppler components that emerge in the multipath fading CSI response which differ at varying frequencies. To further address the static clutter response of the environment, a clutter profile $C[k]$ is computed as the temporal mean of each subcarrier over the observation window. Subtracting this profile from the interpolated CSI suppresses the stationary background, effectively isolating the dynamic variations $H_D(n,k)$ caused by the target motion
\begin{align}
    C[k] 				 &= \frac{1}{N} \sum_{n=0}^{N-1} H[n,k]e^{-j \phi_{\text{ref}}[n]} \\
    H_D[n, k] &= H[n, k] - C[k]
\end{align}

\subsection{Velocity Extraction}

\begin{figure}[t!]
    \centering
    \begin{tabular}{c}
         \includegraphics[width=0.9\linewidth]{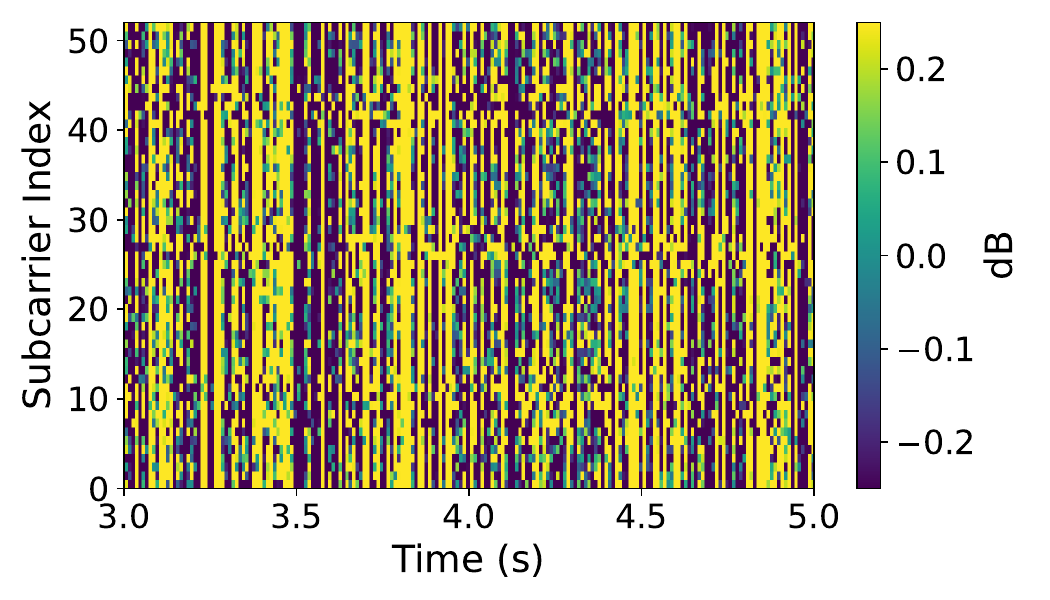} \\ (a) \\
         \includegraphics[width=0.9\linewidth]{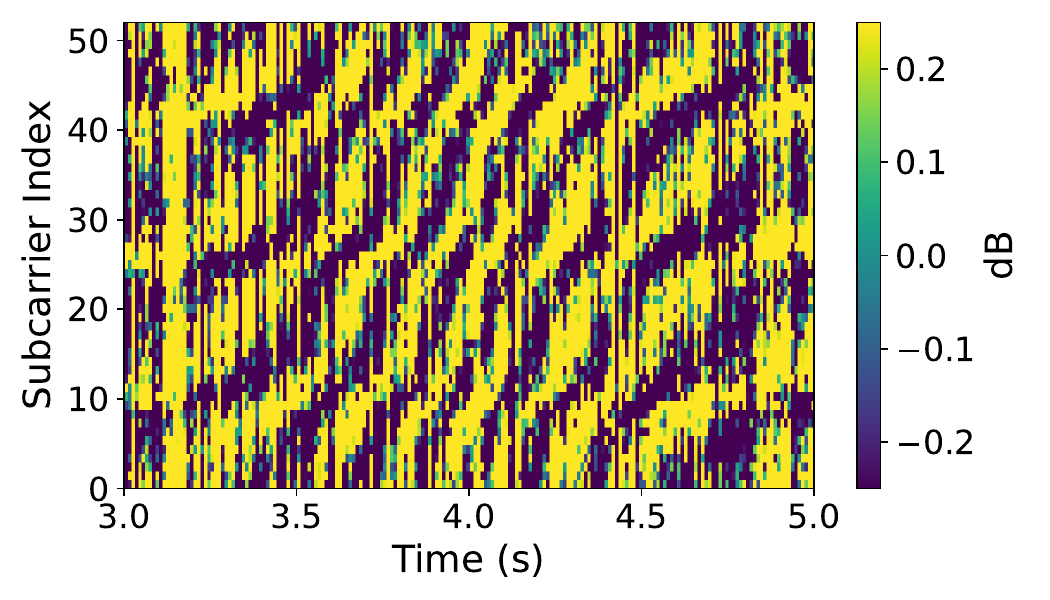} \\ (b) \\
         \includegraphics[width=0.9\linewidth]{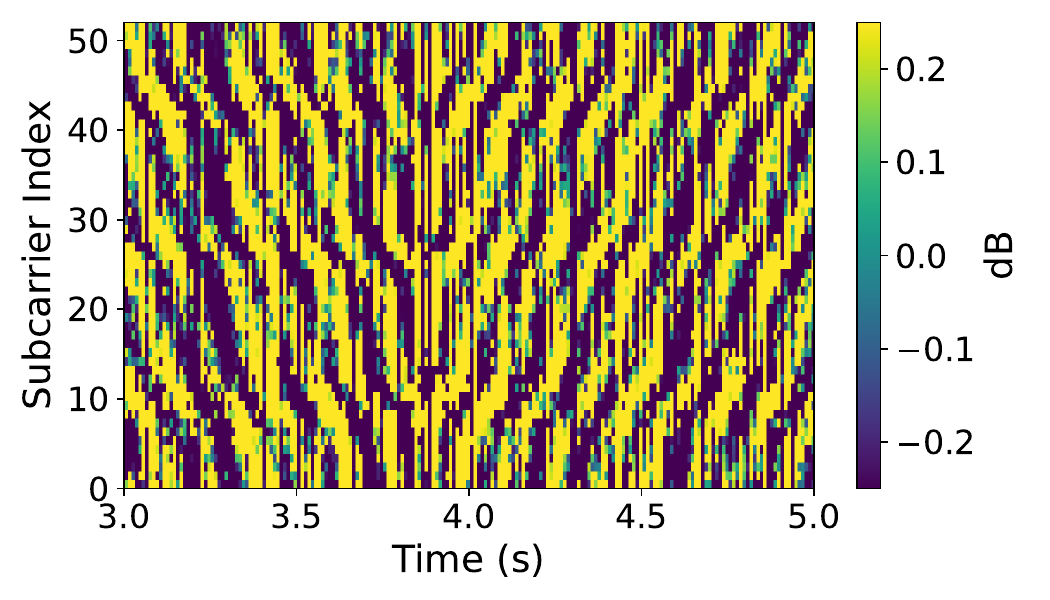} \\ (c)
    \end{tabular}
    \caption{Raw time-frequency representations of the CSI subcarriers for varying motion types, illustrating the impact of (a) static conditions, (b) rapid dynamic variations, and (c) square motion profiles on the channel frequency response.}
    \label{fig: CSI Amplitude}
\end{figure}

Due to frequency-selective fading, target motion affects each subcarrier differently. Subcarriers situated in the nulls of the CSI magnitude response exhibit the highest sensitivity to minute spatial displacements. To visualize this, it becomes practical to consider the amplitude matrix $|H_{{D}}[n, k]|$. This is shown for a simulated channel in Fig.~\ref{fig: CSI Amplitude} for a static scene, one with triangular dynamic motion, and one with dynamic motion in a square wave profile. The simulations modeled motion with a maximum velocity of 0.45 m/s for objects at a range of 0.75 m with an angle of $18^\circ$ between motion and radar boresight. The SNR was set to 5 dB to mimic high noise scenarios. 
The static scene exhibits only noise-like variations, however the two dynamic scenes show clear time-varying amplitude variations in proportion to the motion.

The Doppler frequency of a signal can be calculated by the frequency-scaled time derivative of the phase term $\delta\phi/\delta t$ for any subcarrier. However, due to noise, the ability to estimate Doppler from a single subcarrier is challenging. 
To address this shortcoming, we implement an averaging approach  to extract the Doppler shifts across eight subcarriers that exhibit the largest frequency variation and thus contain the most Doppler information, which are selected by a channel weighting mask $K_{o}$. 
The filtered channel response is then
\begin{equation}
    H_{F}[n,k] = \begin{cases} 
    \frac{\sum_{m \in M} K_{o}[n, k] H_D[n, k]}{\sum_{k \in K} K_{o}[n, k]}, & \text{if } \sum_{m \in M} K_{o}[n, k] > 0 \\ 
    0, & \text{otherwise} 
    \end{cases}
\end{equation}

To further exploit the frequency diversity of the OFDM bandwidth, Principal Component Analysis (PCA) is applied to the dynamic CSI. By treating each subcarrier as a single feature, PCA identifies the orthogonal projection that captures the maximum variance, which in the case of CSI corresponds directly to the movement of targets within the scene~\cite{zollanvari2023machine,Duda2001}. Letting $H_I = \Re(H_F)^T$ and $H_Q = \Im(H_F)^T$, the primary principal components for the real and imaginary parts are given as
\begin{equation}
    h_\text{I} = \text{PCA}_{1}(H_I), \quad h_\text{Q} = \text{PCA}_{1}(H_Q)
\end{equation}
such that
\begin{equation}
   h_\text{D}[n] = h_I[n] + j h_Q[n]
\end{equation}

During dynamic motion of the transponder, the received signal often experiences significant amplitude fading, which can introduce additional error in the received Doppler waveform. To counter this, an amplitude invariant approach is implemented using a binary comparator. For an arbitrary input sequence $x[n] \in \{\Re(H_\text{D}[n]), \Im(H_\text{D}[n])\}$, the comparator utilizes hysteresis bounds $[A_\text{high}, -A_\text{low}]$ to generate a discrete state output $H_\text{D}[n]$
\begin{equation}
    H_\text{D}[n] = \begin{cases} 
    1, & \text{if } x[n] \ge A_\text{high} \\ 
    -1, & \text{if } x[n] \le A_\text{low} \\ 
    H[n-1], & \text{otherwise} 
    \end{cases}
\end{equation}
This operation effectively acts as a hard limiter, stripping away amplitude distortion and focusing purely on the signal phase progression where $A_\text{low}$ and $A_\text{high}$ are set to $10\%$ of the standard deviation of the dominant PCA output such that frequency content of the signal is preserved. The resulting in-phase $I$ and quadrature $Q$ components are then interpolated back into a sinusoid and the mean subtracted to achieve an interpolated Doppler response and mitigate residual dc bias. The interpolation is performed using an $8^\text{th}$ order Savitzky-Golay filter with a window length of 25, yielding an interpolated complex baseband signal

\begin{equation}
    s_c[n] = \text{SavGol}(I_{\text{bin}}[n], 25, 8) + j \, \text{SavGol}(Q_{\text{bin}}[n], 25, 8)
\end{equation}

\begin{figure}[t!]
    \centering
    \includegraphics[width=\linewidth]{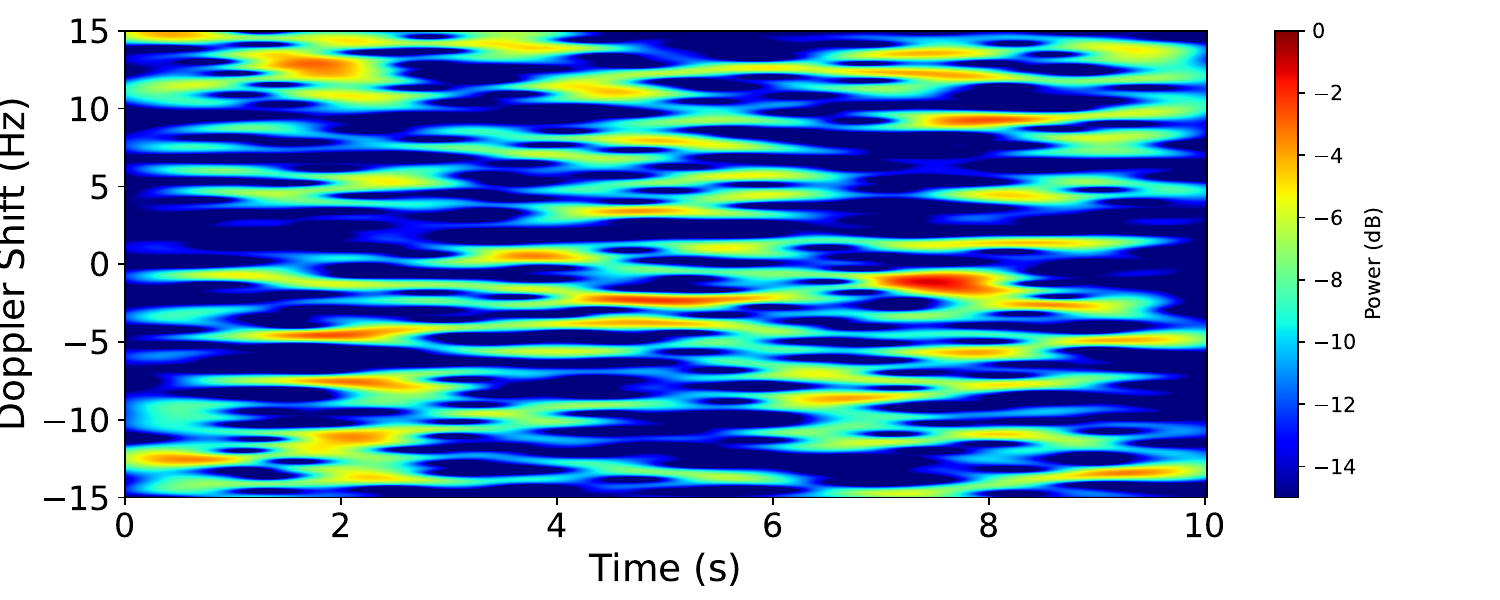} \\
		(a)
		\\
    \includegraphics[width=\linewidth]{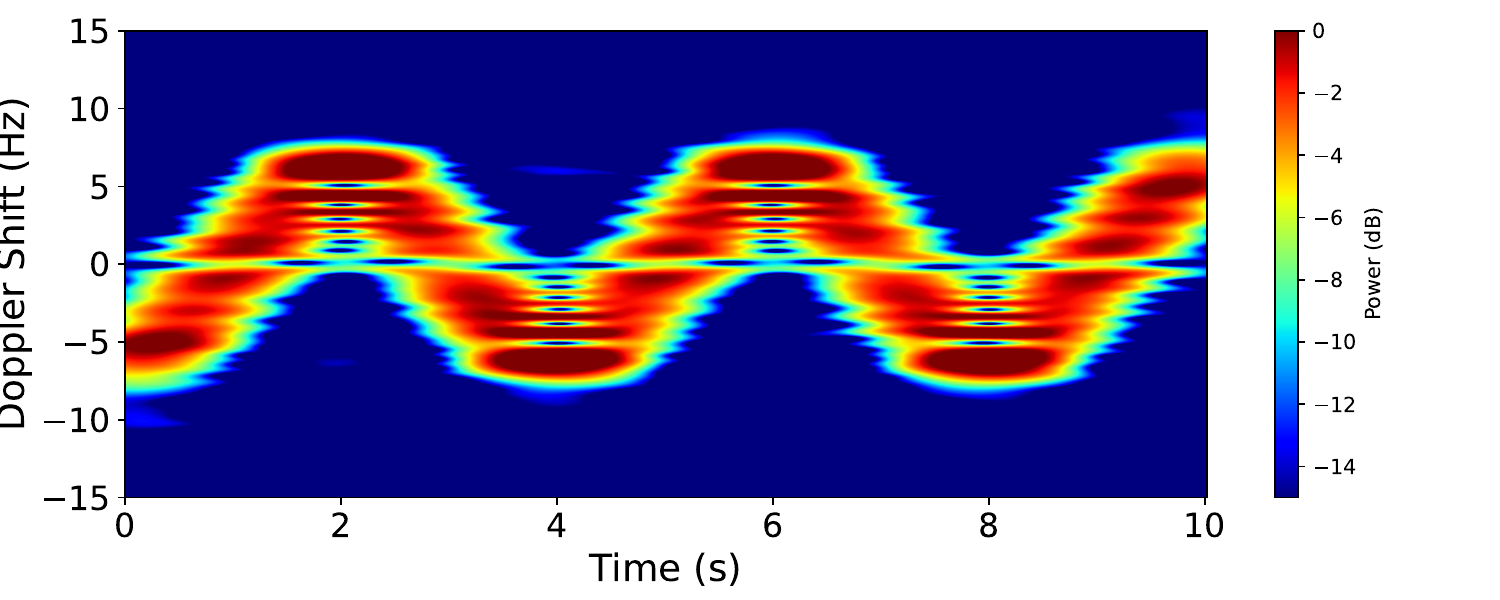} \\
		(b)
    \caption{Time-frequency representation of the CSI Doppler signatures. (a) The spectrogram derived from raw CSI, exhibiting severe noise and clutter masking the true target kinematics. (b) The enhanced spectrogram post-processing, revealing the clear micro-Doppler signature of the target.}
    \label{fig: Simulated CSI Spectrogram}
\end{figure}

To visualize the target's temporal velocity profile, a Short-Time Fourier Transform (STFT) is applied to $s_c[n]$. The time-frequency spectrum of a single subcarrier for the raw simulated channel models response is shown in Fig.~\ref{fig: CSI Amplitude}(a), while the mitigated response after interpolation of the PCA is shown in Fig.~\ref{fig: Simulated CSI Spectrogram}(b), showing a clear improvement in the triangular motion estimation. 

\section{Transponder and Wireless System Design}
\label{sec: System}

\subsection{Passive Nonlinear Transponder Design}

\begin{figure}[t!]
    \centering
    \includegraphics[width=0.75\linewidth]{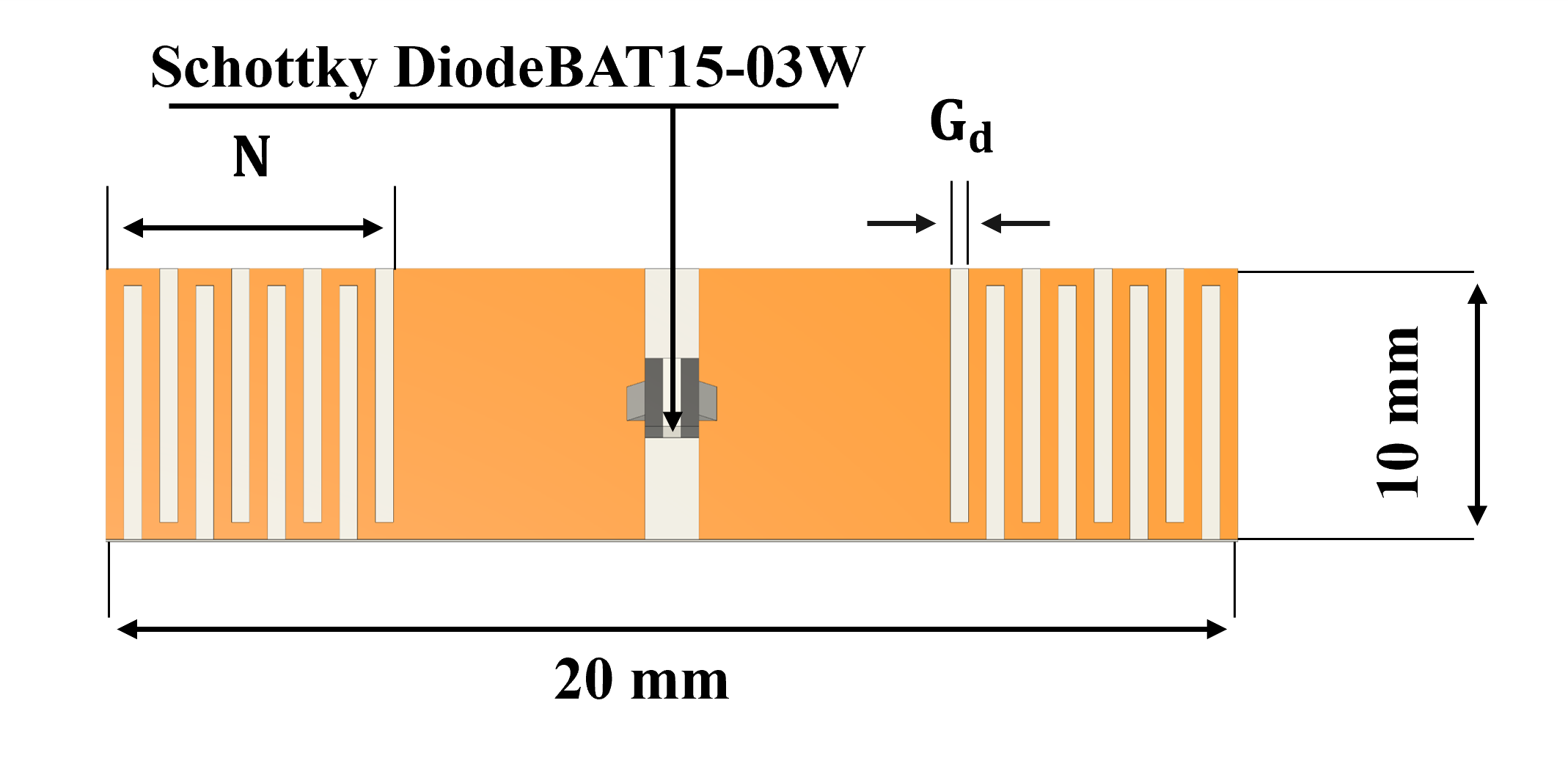}\\ (a) \\
    \includegraphics[width=0.5\linewidth]{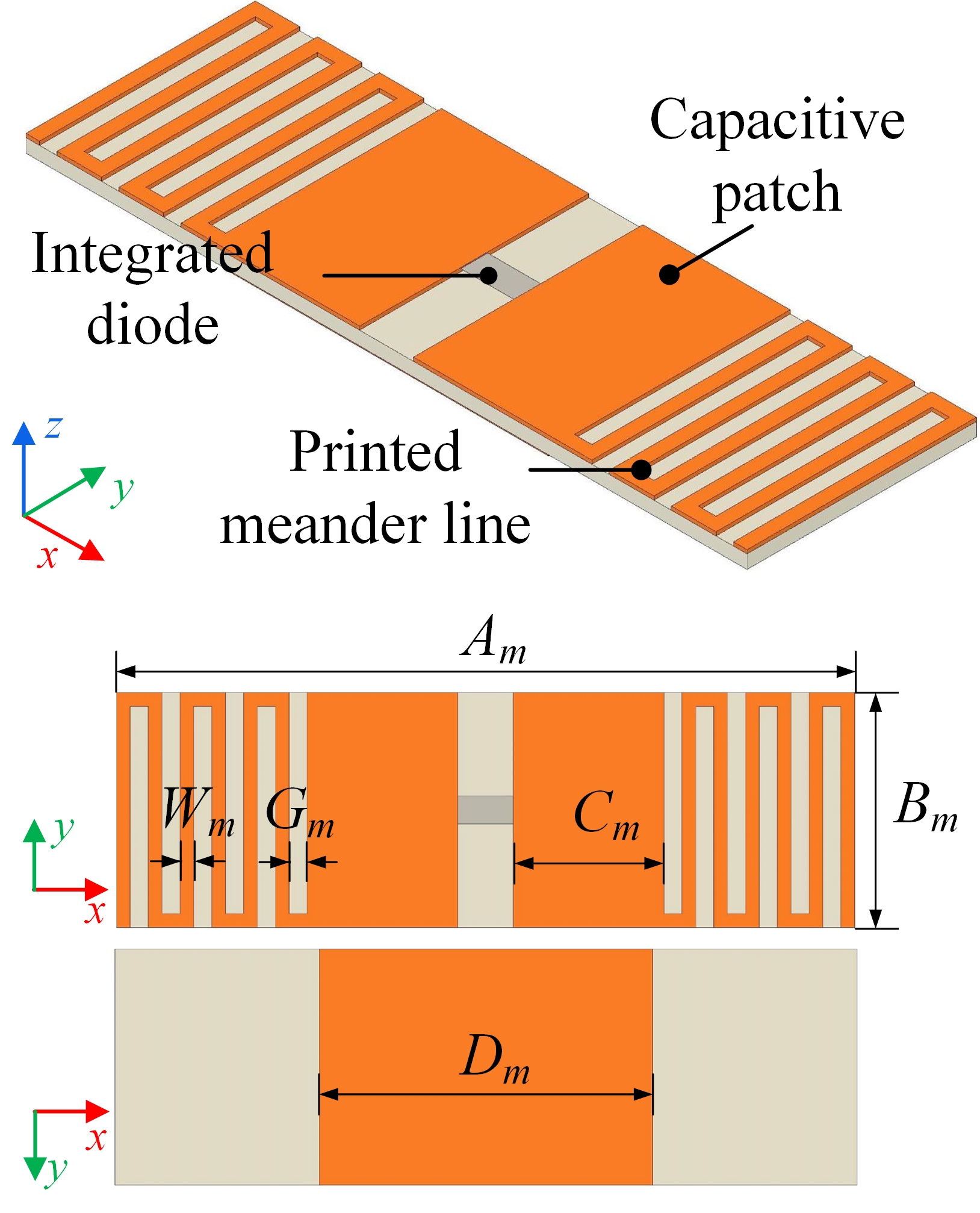}\\ (b)
    \caption{Schematic and geometry of the passive nonlinear transponder. (a) Top view detailing the integrated BAT15-03W Schottky diode, meander lines, and capacitive patches. (b) 3D and dimensional breakdown of the printed monopole structure operating at 2.44 GHz.}
    \label{fig: transponder}
\end{figure}

To ensure that the nonlinearities are sufficiently retransmitted to the system, a small passive transponder operating at 2.4~GHz was designed based on the design in~\cite{ShengRFFingerprinting,Hilton_CAMA}. A mirrored inverted top hat monopole was designed on a single layer of 0.5 mm Rogers 4350b substrate and mirrored to operate in an quasi-dipole layout. The port placed between the two elements was a nonlinear device chosen to be the BAT15-03W Schottky diode as displayed in Fig.~\ref{fig: transponder}. 
The resonance frequency was achieved by heuristically sweeping the electrical length of the element until it was tuned to Wi-Fi frequencies. This is achieved by modifying the $N$ number of turns on the microstrip and the inter-arm gap, $G_d$ which modifies the self-reactance between arms of the element. An additional ground layer was utilized for to increase capacitance, further reducing the physical size of the device while improving impedance matching to the diode via the self-resonance of the capacitive top-hat patches counteracting that of the monopoles. This results in a compact design with a $ka =0.56$, where $k$ is the wavenumber and $a$ is the maximum dimension of the antenna. The simulated transponder achieved a realized gain of 1.9 dBi in a 30 MHz bandwidth (measured at -10 dB S11) while maintaining 
omnidirectional radiation in the H-plane with an electrically small planar size of $0.21 \times 0.06\,\lambda^{2}$ with the design parameters shown in Table \ref{tab:table1}. This makes the tag an ideal choice for observing a 20 MHz channel bandwidth in 802.11 applications, theoretically enabling secure device authentication while also performing velocity estimation.

\begin{table}[t]
\caption{Antenna Dimensions \label{tab:table1}}
\centering
\renewcommand{\arraystretch}{1.4}
{
\begin{tabular}{||c|c|c|c|c|c|c||} \hline 
Parameter & \(A_{m}\)&\(B_{m}\) &\(C_{m}\) & \(D_{m}\) &\(G_{m}\) & \(W_{m}\)\\  \hline
Value (mm) & 28 & 8.2 & 6 & 12 & 0.8 & 0.5\\\hline 
\end{tabular}
}
\end{table}

\subsection{CSI micro-Doppler Radar Design}

\begin{figure}

    \centering
    \includegraphics[width=0.75\linewidth]{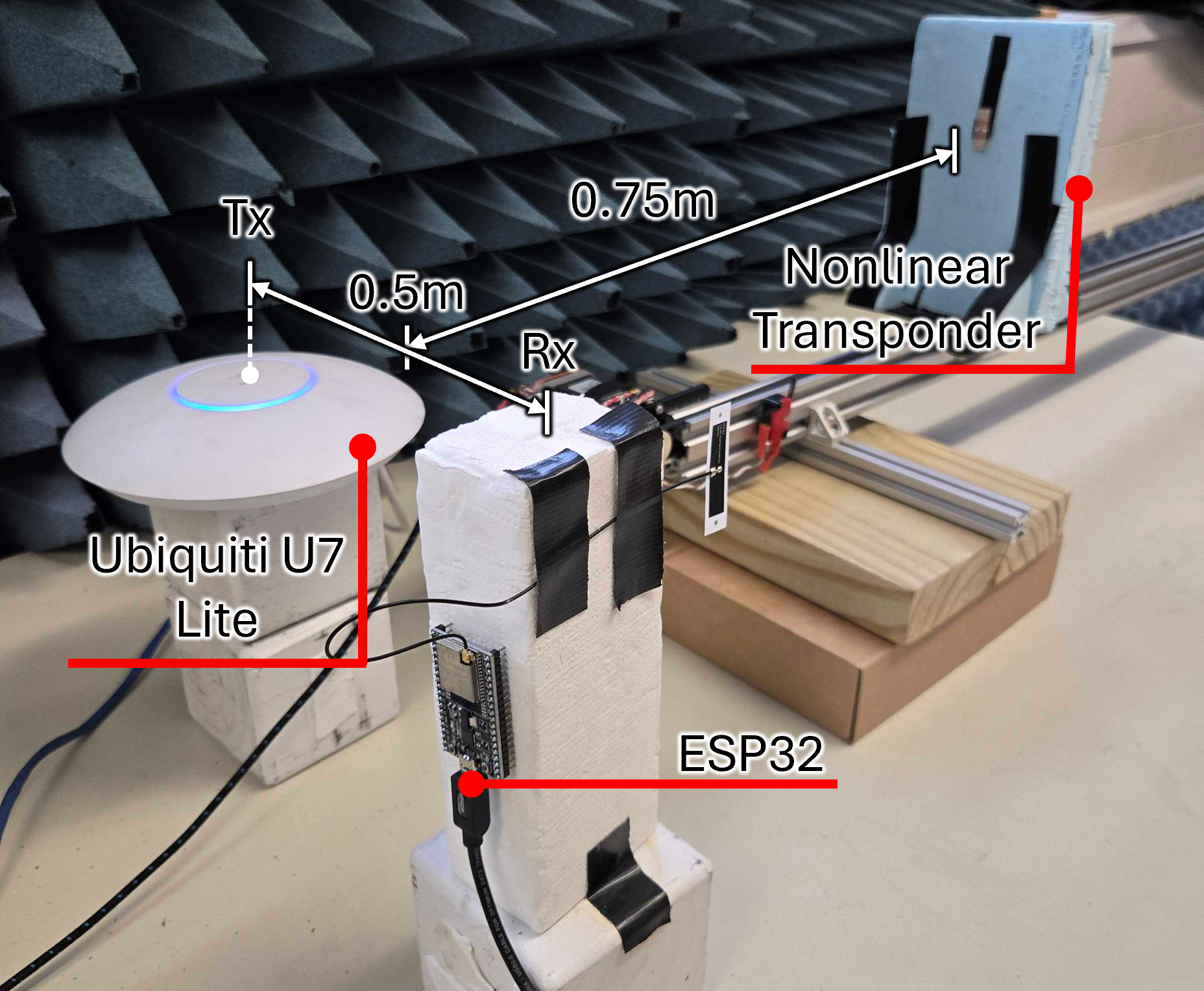}
    \caption{Experimental setup in a semi-anechoic environment, featuring the Ubiquiti U7 AP transmitter, the ESP32 receiver, and the passive nonlinear transponder mounted on a linear actuator for precise dynamic motion profiling.}
    \label{fig: Setup}

\end{figure}

The wireless system is based on the Ubiquiti U7 AP as the transmitter and the ESP32 as the receiver each operating on 802.11 channel 11 at 2.42 GHz. Each device was chosen for the high adoption rate, low-cost, and the ability to access channel state information. Primarily, to ensure that the nonlinearities of the transponder are dominant, and thus that IMPs are retransmitted, the incident power at the nonlinear device should be high enough to forward-bias the device. It has been demonstrated in previous works~\cite{ShengRFFingerprinting,11075563,colpitts2004harmonic} that the device must be biased within the nonlinear or exponential region of the I-V curve to generate sufficient nonlinearities for classification. The power requirements can then be estimated based on the maximum range in a noiseless radar system, given by
\begin{equation}
R = \left[ \frac{P_t G_{t} G_{r}\lambda^2 \sigma_t }{P_r (4\pi)^3} \right]^{\frac{1}{4}}
\end{equation}
where $P_t$ is the transmit power, $P_r$ is the received power, $G_t$ and $G_r$ are the gains of the radar transmit and receive antennas, $\lambda$ is the wavelength, and $\sigma_t$ is the equivalent scattering cross section of the transponder, given by
\begin{equation}
\sigma_t = \epsilon_c \frac{\lambda}{4\pi}G_{t1} G_{t2}
\end{equation}
where $\epsilon_c$ is the conversion efficiency of the transponder and $G_{t1}$ and $G_{t2}$ are the gains of the two transponder antennas. For all antenna gains equal to 2 dBi, a transmit power of 20 dBm enables a maximum read-range of $\approx$ 4.5 m.

\begin{figure}[t!] 
    \centering
    \includegraphics[width=\linewidth]{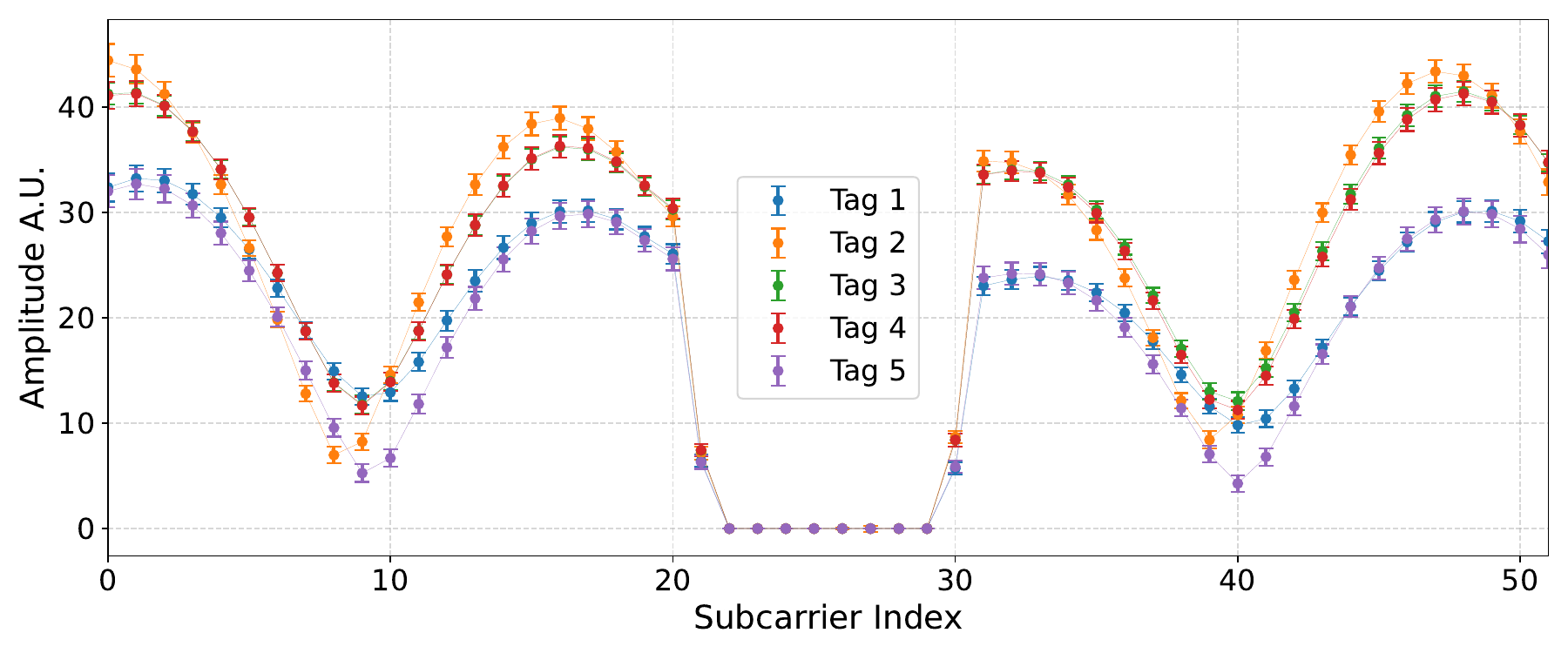}\\ (a) \\
    \includegraphics[width=\linewidth]{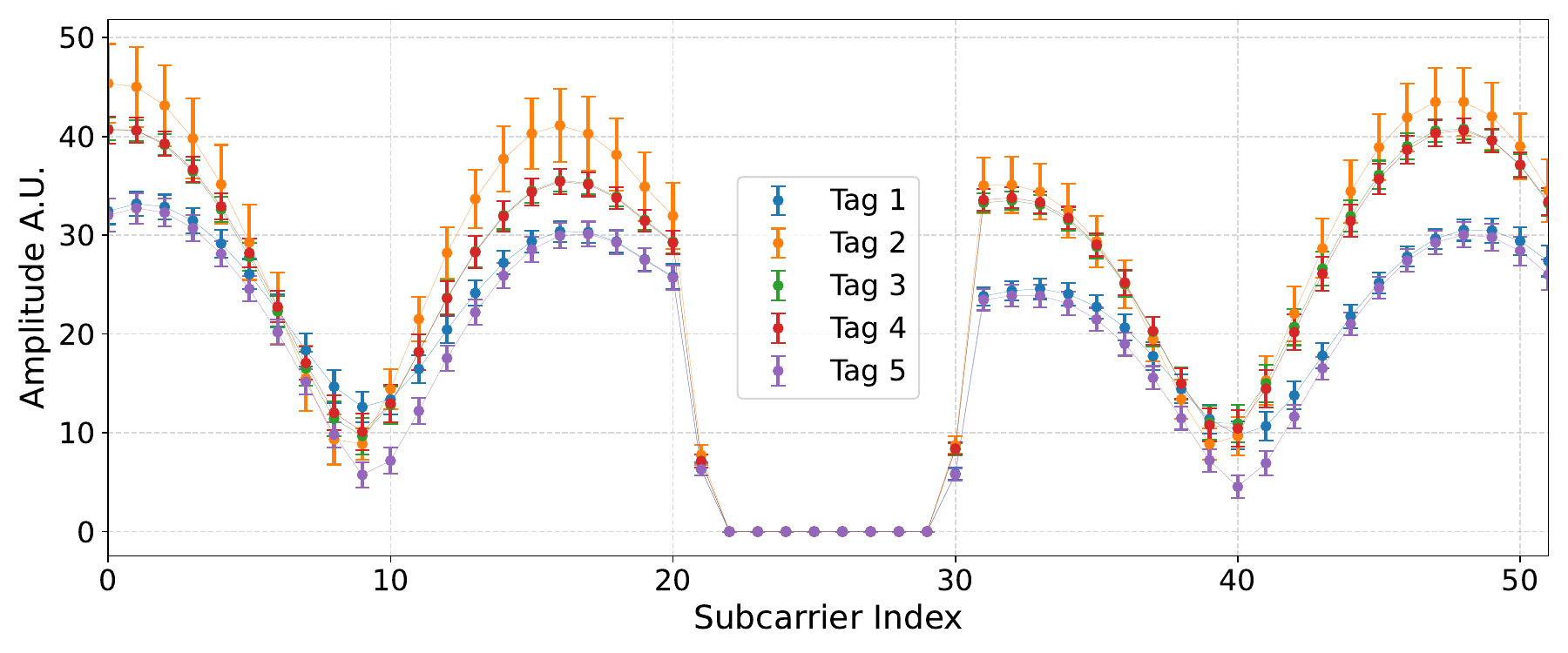}\\ (b) \\
    \includegraphics[width=\linewidth]{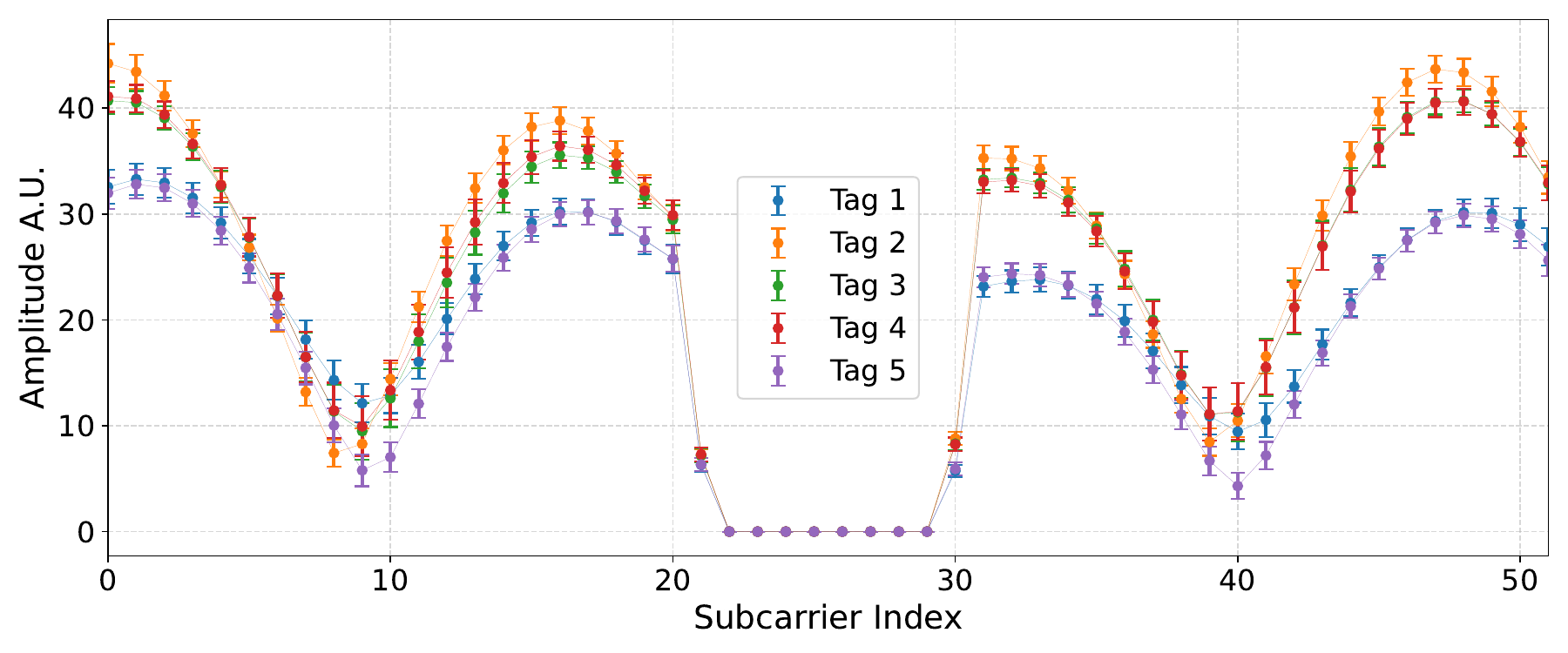}\\ (c) \\
    \caption{Measured CSI magnitude responses across 52 subcarriers for five distinct passive tags (Tag 1–5) in the case of (a) static, (b) constant acceleration (triangular velocity profile), (c) constant velocity (square-wave velocity profile). These each demonstrate the different spectral signatures and amplitude variations inherent to each device.}
    \label{fig: CSI Fingerprint}
\end{figure}

\section{Measurements}
\label{sec: Measurement}


\begin{figure}[t!]
    \centering
    \includegraphics[width=\linewidth]{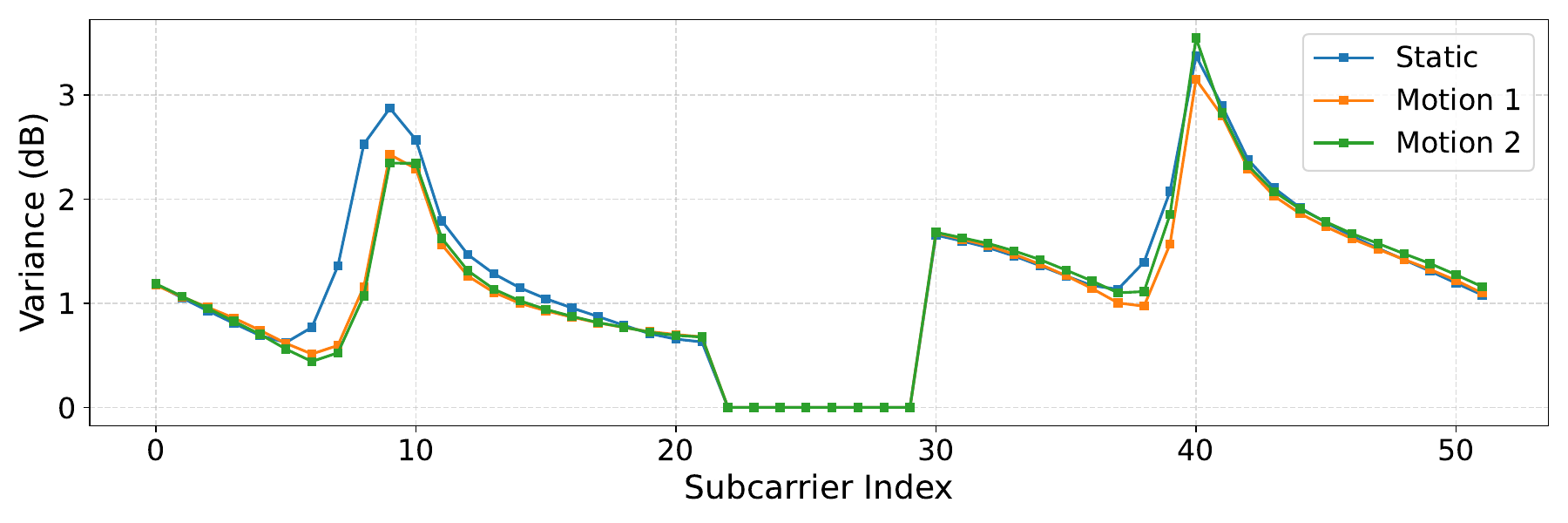}
    \caption{Measured CSI magnitude responses across 52 subcarriers for five distinct passive tags (Tag 1–5), demonstrating the unique spectral signatures and amplitude variations inherent to each device. (a) Static, (b) constant acceleration (triangular velocity profile), (c) constant velocity (square-wave velocity profile).}
    \label{fig: CSI Var}
\end{figure}


To verify the RF fingerprinting performance of the tag, the system was first set up in a static configuration as shown in Fig.~\ref{fig: Setup}. The AP and ESP32 were placed 0.25 m equidistant from the mounting plate of a linear actuator that was used to determine the precise down-range position of the transponder. For this work, five tags were manufactured and tested, all operating at 2.42 GHz to ensure the stability of the estimation approach. The nonlinear RF fingerprinting measurements were then conducted by illuminating the passive transponders with L-LTF waveforms generated by the Uniquiti AP at a range of 0.75 m. The ESP32 receiver captured the backscattered signals, which contained the unique IMPs generated by the Schottky diode. 
The motion of the transponder was evaluated by mounting the transponder on a linear actuator with position error $\leq$1 mm as shown in Fig.~\ref{fig: Setup}. 
The time-varying frequency shift of the transponder signal was  characterized by the relative magnitude change in the across the channel-dependent subcarriers in the channel state information (CSI) as the transponder moved along two defined motion profiles over a range of 0.25 m to 0.75 m: constant velocity and constant acceleration, each with a maximum ~ 0.5 m/s. Samples were then captured continuously from the ESP32 in batches of 1000 packets at a sample rate of 100 Sa/s. 

The results of the three motion profiles are shown in Fig.~\ref{fig: CSI Fingerprint}, where it can be seen that due to the high SNR of the tag and quasi-static environment, each tag retains a unique spectral response by observing the CSI amplitude across all subcarriers. The measures CSI responses are in good agreement with the simulation of Fig.~\ref{fig: Sim CSI Fingerprint}. The measured responses demonstrate an average standard deviation of $\approx$1.15 dB across all CSI channels, as shown in Fig.~\ref{fig: CSI Var}. 
Peaks in the variance correspond to nulls in the CSI, where the SNR is low.

\begin{figure}[t!]
    \centering
    \begin{tabular}{c}
    \includegraphics[width=0.9\linewidth]{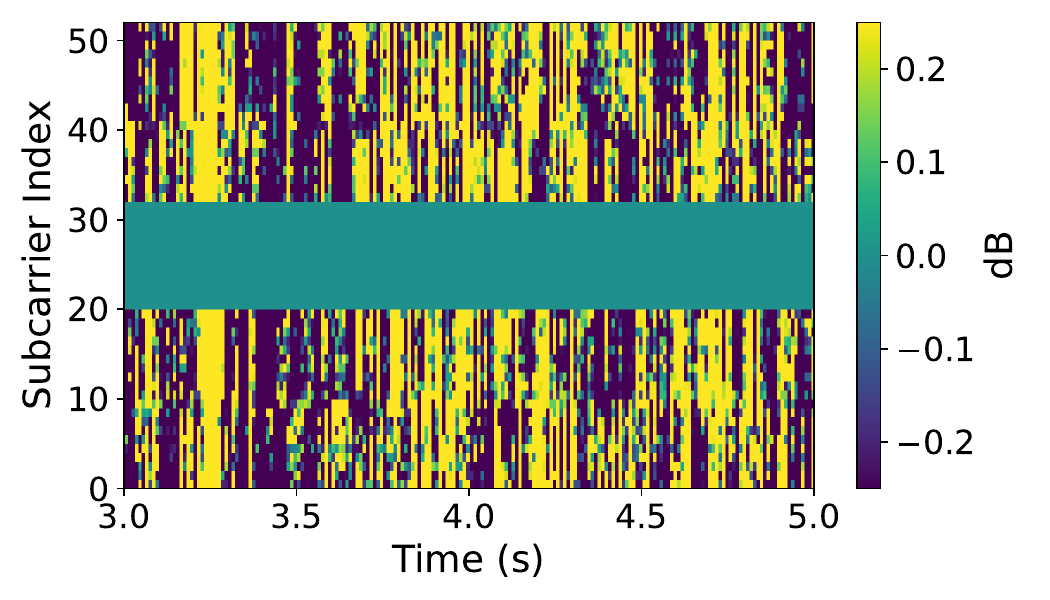}\\ (a) \\
    \includegraphics[width=0.9\linewidth]{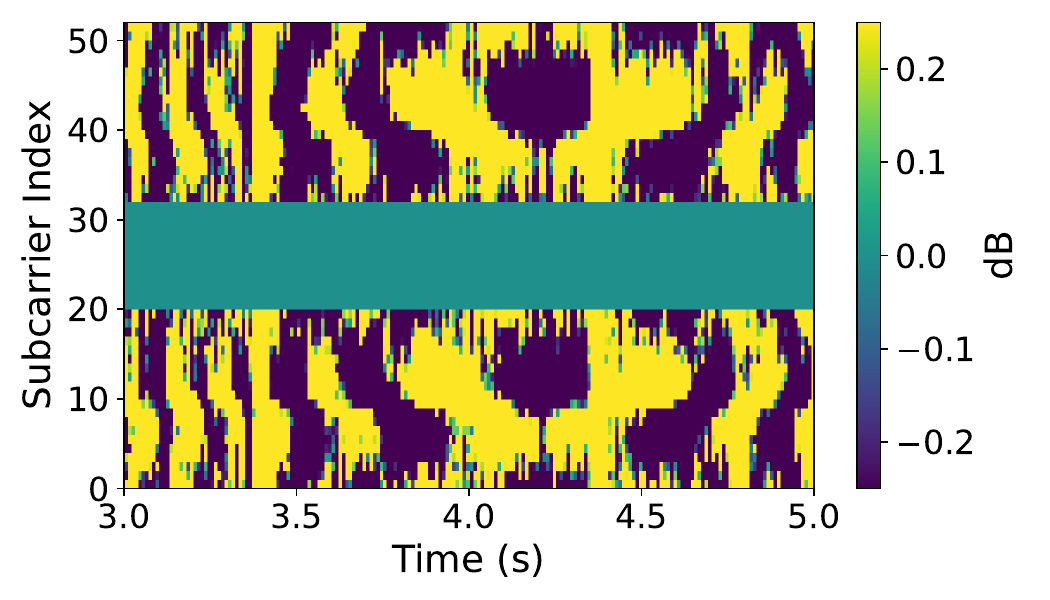}\\ (b) \\
    \includegraphics[width=0.9\linewidth]{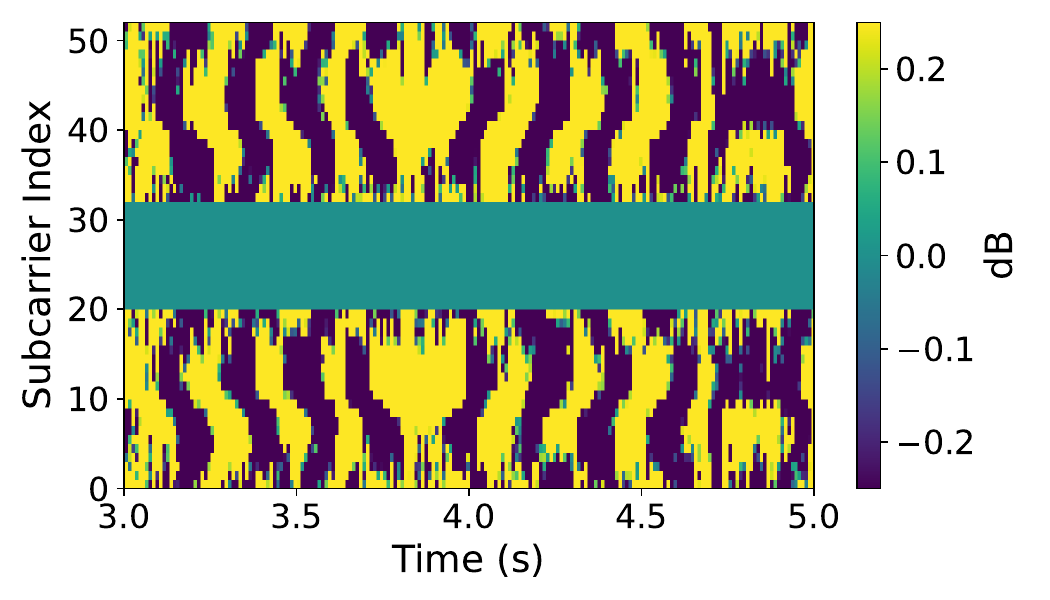}\\ (c)
    \end{tabular}
    \caption{Processed CSI subcarrier index heatmaps across 5-second and 10-second observation windows, highlighting the distinct temporal variations captured during the linear actuator motion profiles.}
    \label{fig: Meas CSI Amplitude}
\end{figure}

\begin{figure*}[h!]
    \centering
    \begin{tabular}{cc}
    \includegraphics[width=0.5\linewidth]{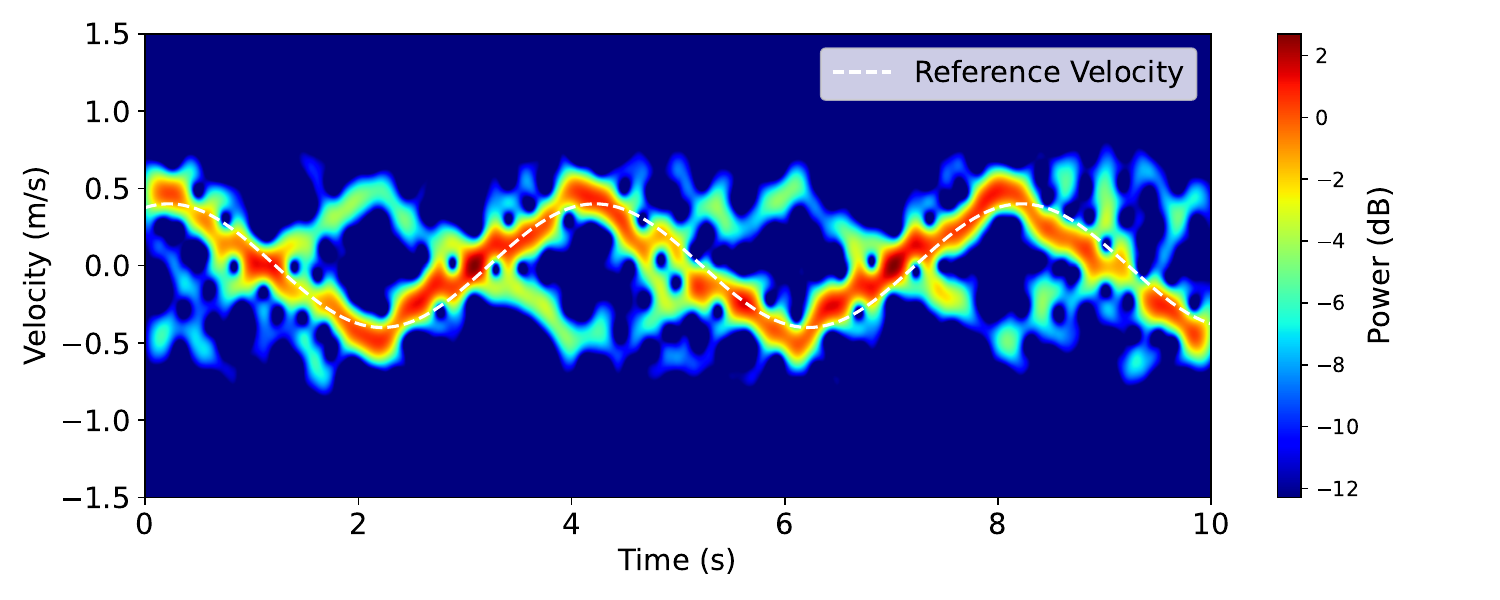} &
    \includegraphics[width=0.5\linewidth]{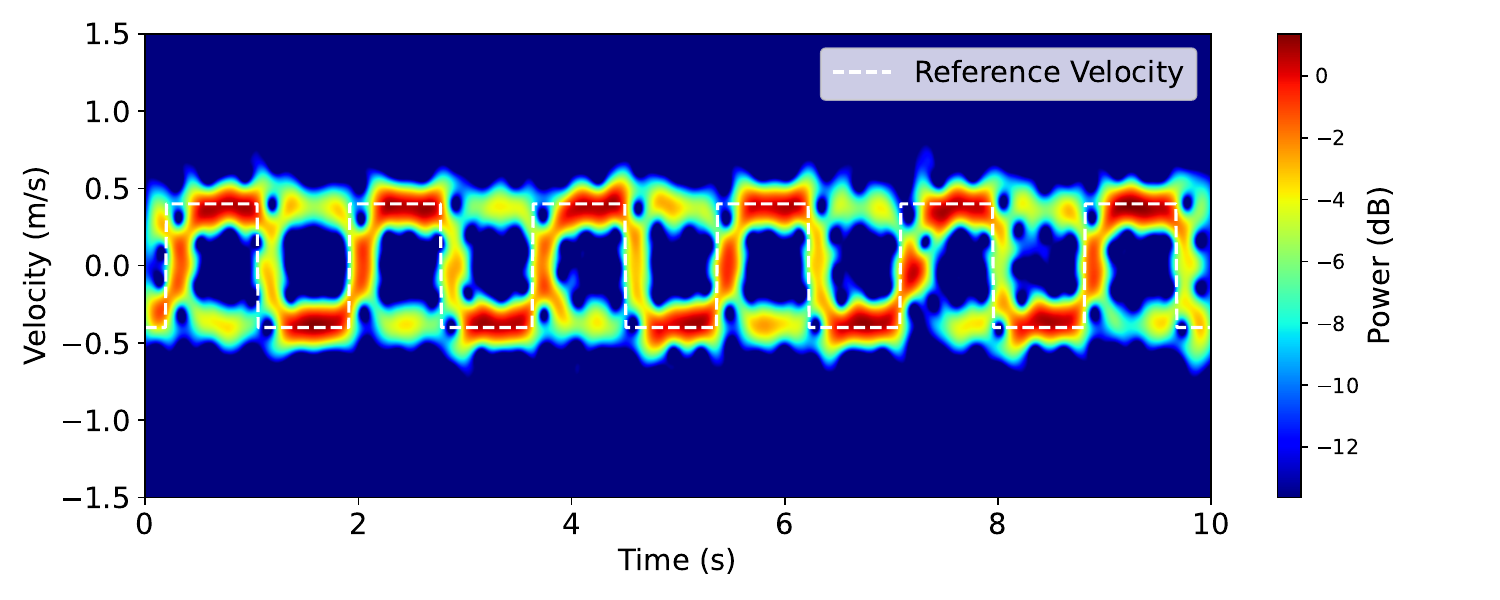} \\ a & b \\
    \includegraphics[width=0.5\linewidth]{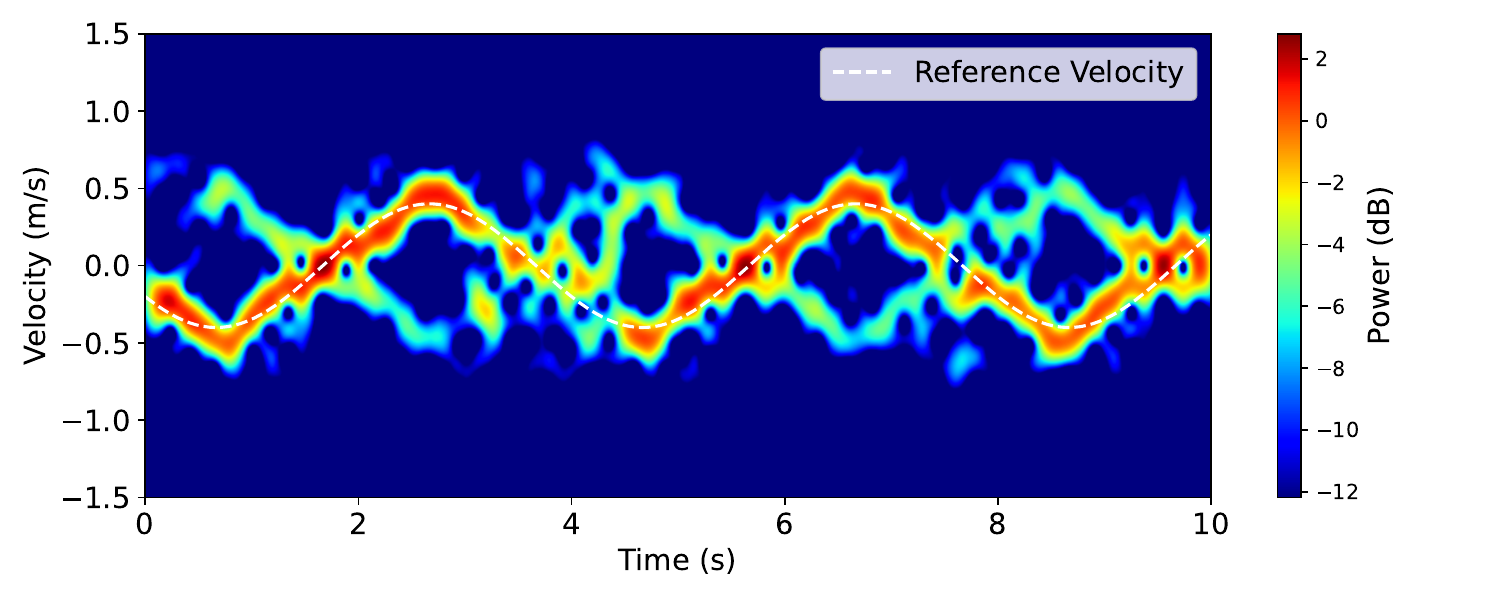} &
    \includegraphics[width=0.5\linewidth]{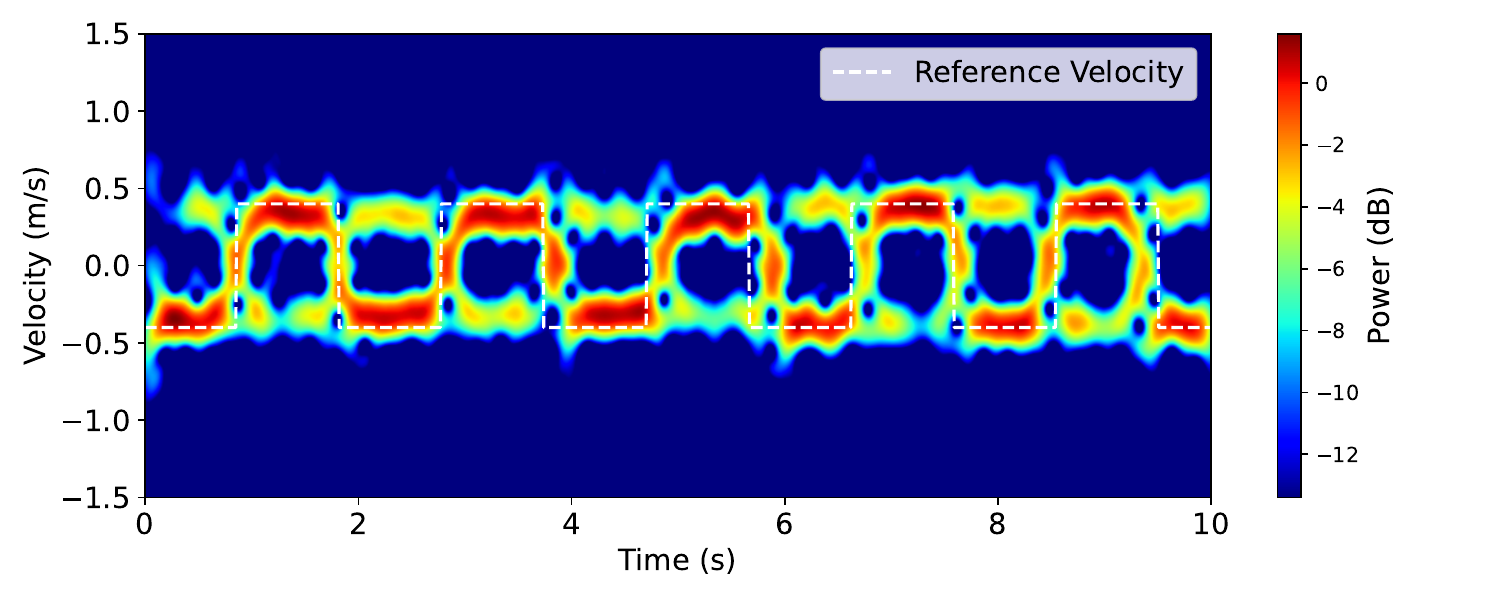} \\ c & d
    \end{tabular}
    \caption{Extracted Doppler shifts showcasing the sanitized micro-Doppler signatures for specific tag and motion combinations for (a) tag 1 with constant acceleration, (b) tag 1 with constant velocity, (c) tag 3 with constant acceleration, and (d) tag 3 with constant velocity. The spectrogram output is achieved after applying the amplitude-invariant binary comparator and filtering.}
    \label{fig: meas spectrograms}
\end{figure*}






The measured results of the CSI amplitude response are shown in Fig.~\ref{fig: Meas CSI Amplitude}, where it is shown that for the cases of a transponder in motion demonstrates good agreement with simulated results in Fig.~\ref{fig: CSI Amplitude}.
Applying the phase calibration and PCA decomposition procedures to the measured subcarrier responses then results in the measured micro-motions for each velocity profile as shown in the spectrograms (magnitudes of the short-time Fourier transform) of Fig.~\ref{fig: meas spectrograms}. The Doppler frequency is given by $f_\text{D} = 2f_cv/c$, where $c$ is the speed of light, $f_c$ is the carrier frequency, and $v$ is the velocity of the transponder. This results in an expected Doppler shift of 6.5 Hz for each target at a carrier of 2.42 GHz and radial velocity of 0.5 m/s. The measured responses agree well with the calculated response of each motion profile, with an root-mean square error of 2.21 cm/s.

\section{Device identification and Gesture Classification}
\label{sec: Classification}

Based on the nonlinear responses of the transponders and the motion profiles recovered from the time-varying CSI data, we show in this section that both device identification and motion classification can be accomplished with high accuracy. We demonstrate device identification using frequency samples from the CSI response, and motion classification from the spectrograms obtained from the time-varying amplitudes of the CSI profiles. These data are input into machine learning algorithms as described below.


\subsection{Device identification} 
    
\begin{figure}[t!]
    \centering
 
    \includegraphics[width=0.75\linewidth]{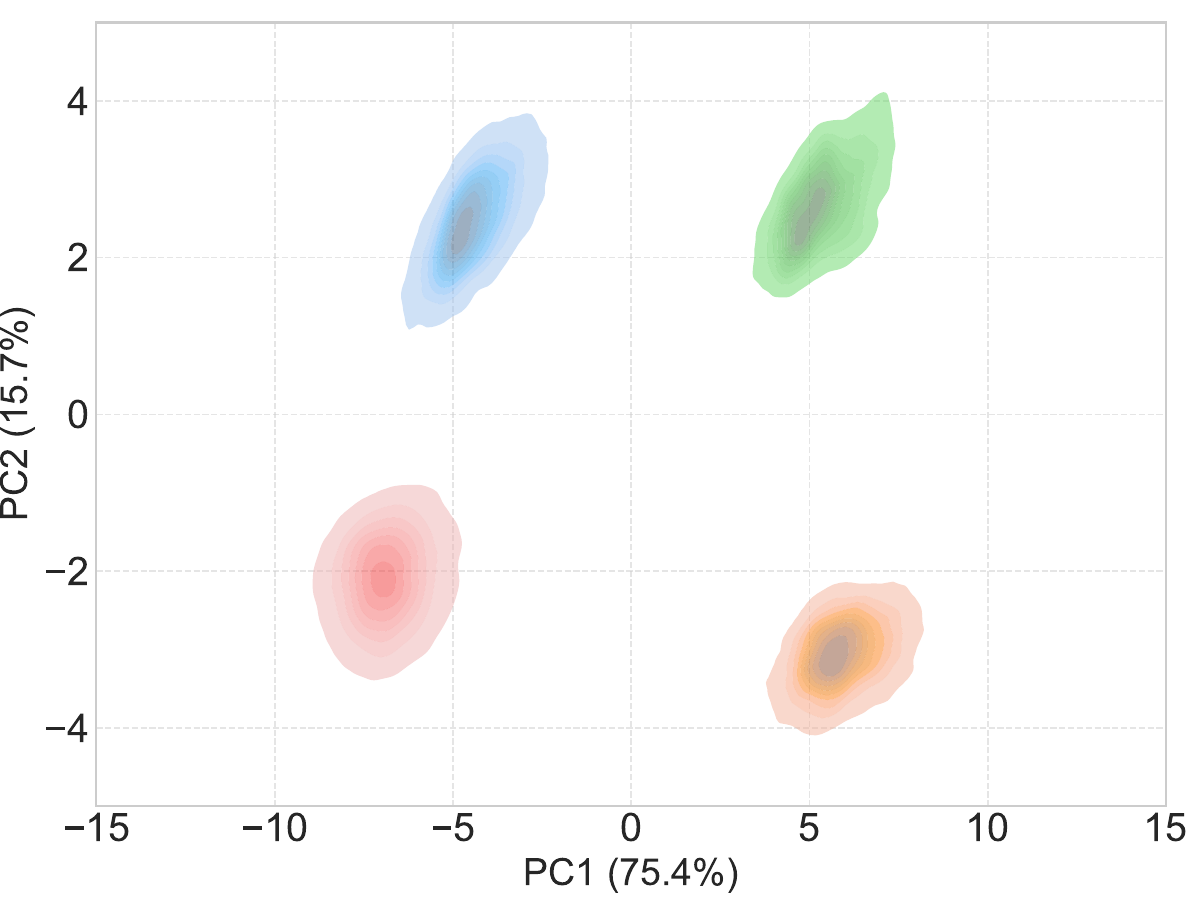} \\ (a) \\ 
    \includegraphics[width=0.75\linewidth]{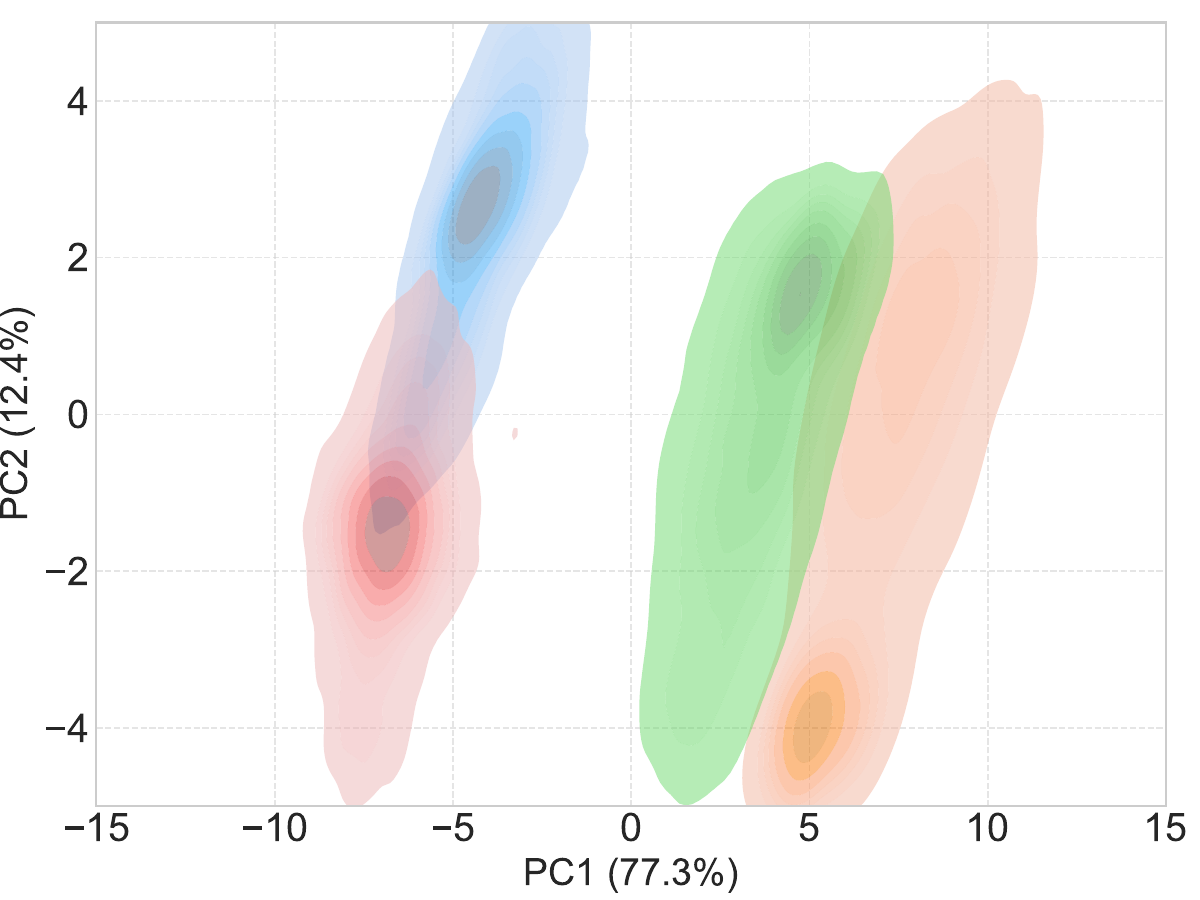} \\ (b) 

    \caption{Principal Component Analysis (PCA) scatter density plots illustrating the clustering and phase evolution of the in-phase ($I$) and quadrature ($Q$) components for six unique classification classes. {(a) Static PCA response of each tested tag, (b) Dynamic PCA response of each tested tag}. The percentages denote the variance ratios for each principal component.}
    \label{fig: pca}
\end{figure}

\begin{figure}[t!]
    \centering
    \includegraphics[width=0.8\linewidth]{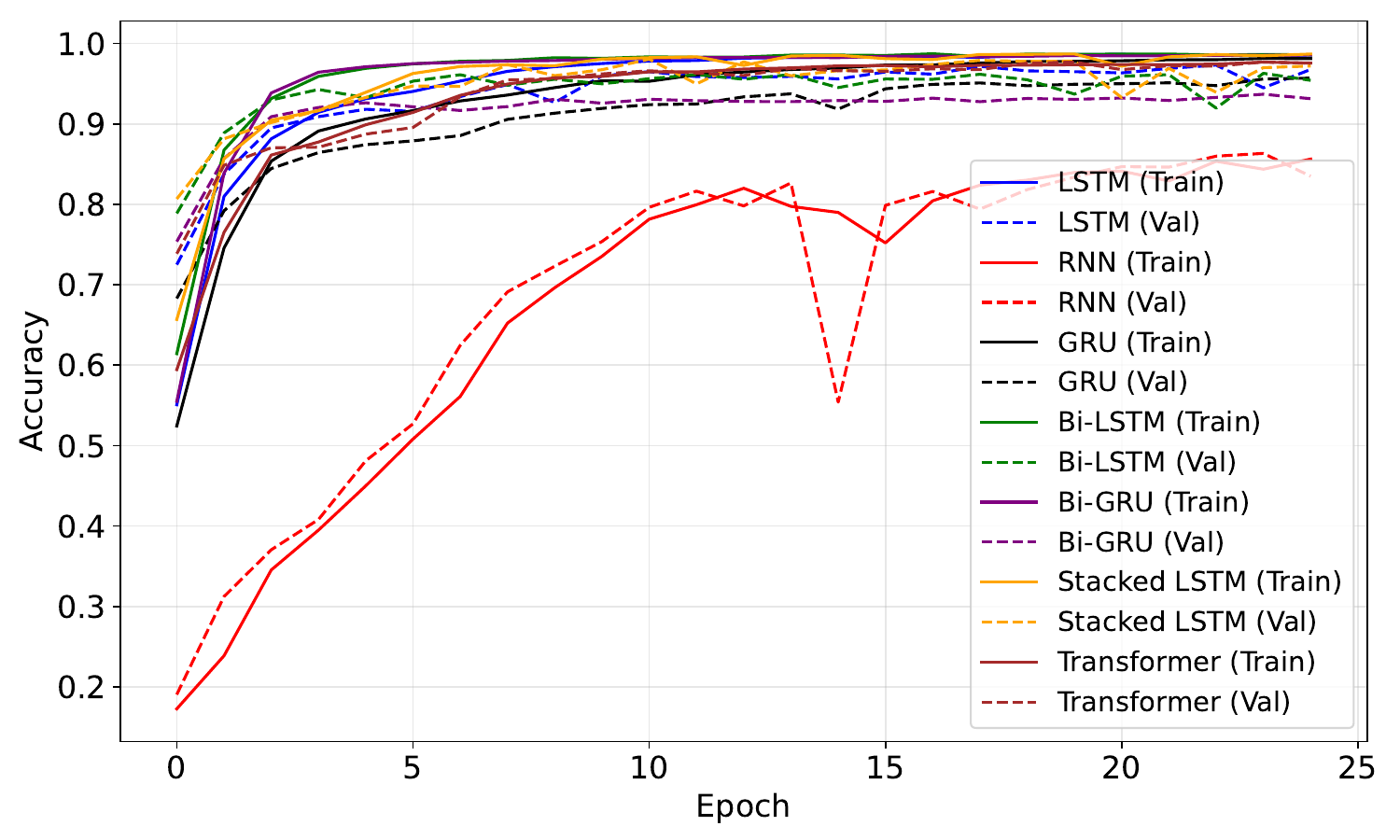} \\ (a) \\ 
    \includegraphics[width=0.8\linewidth]{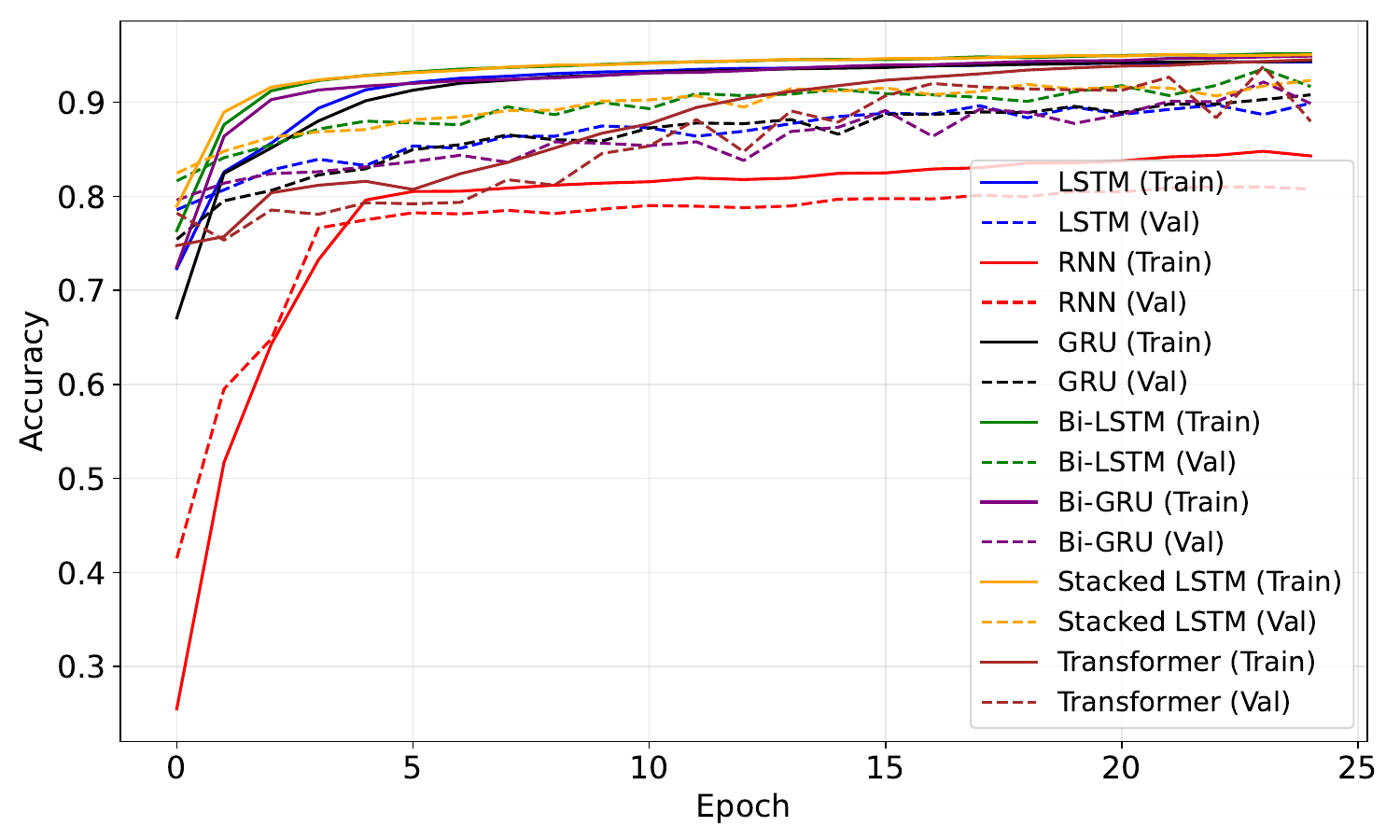} \\ (b)
    \caption{Train and validation accuracy curves for the machine learning models for (a) the static dataset and (b) static and dynamic dataset over 25 epochs across various architectures (LSTM, RNN, GRU, Transformer), demonstrating high convergence and classification accuracy with minimal overtraining.}
    \label{fig: accuracies}
\end{figure}

\begin{table*}[t!]
        \centering
        \caption{\textbf{ Training classifiers on the static case only}: 
				Classification accuracies for the static case (no perturbation), the small 
				perturbation, and the large perturbation cases.}
        \label{tab:static_training}
        \begin{tabular}{||c||c|c|c|c|c|c|c||}\hline
           \textbf{Case} & \textbf{LSTM} & \textbf{RNN}& \textbf{GRU} & \textbf{Bi-LSTM} & \textbf{Bi-GRU} & \textbf{Stacked LSTM} & \textbf{Transformer}\\ \hline
		    Static Case (Train) & 99.9\% & 99.9\% & 99.9\% & 99.9\% & 99.9\% & 99.9\% & 99.9 \%\\ \hline
            Static Case (Val) & 99.9\% & 99.9\% & 99.9\% & 99.9\% & 99.9\% & 99.9\% & 99.9 \%\\ \hline
           Dynamic Case (Train) & 94.3\% & 84.3\% & 94.4\% & 95.2 \% & 94.9 \% &  95.1\% & 94.5\% \\ \hline	
           Dynamic Case (Val) & 89.9\% & 80.8\% & 90.1\% & 91.7 \% & 89.9 \% &  92.3\% & 91.4\% \\ \hline	
           					
        \end{tabular}
\end{table*}
		
To reduce the dimensionality of the feature space and limit the model complexity, we select only a single set of subcarriers as an input to the models. The CSI information was obtained using the methods discussed in Section \ref{sec: Theory Diode}, where approximately 25 minutes ($150\times10^3$ samples) were collected for each of the five transponders. The magnitude of the CSI response was used to mitigate the impacts of phase errors and noise. 
To further mitigate transient signals and thermal noise and to enhance the effective signal-to-noise ratio (SNR), the magnitude responses at each time instance were averaged across three acquisition windows per time sample. The feature vector was then chosen to map to each discrete frequency bin with a tone separation of 312 kHz within the 20 MHz channel bandwidth. Principal component analysis (PCA) was employed to project the correlated subcarrier responses onto a lower-dimensional basis capturing the dominant variation associated the unique spectral response of each target. The results of the PCA space {for the static tags} are shown in Fig.~\ref{fig: pca}(a), {while the dynamic response is shown  in Fig.~\ref{fig: pca}(b).} 

The information from all captured and processed frames were then labeled for supervised training with a standard 80/20 test-train split before being input into a series of time-dependent machine learning models to demonstrate device identification. The models chosen required a level of time invariance, enabling sufficient robustness to temporal and thermal variation while retaining the identification of the unique tag spectral amplitude ratios. Thus, a series of recurrent neural networks were implemented with similar parameters to evaluate the dataset~{\cite{zollanvari2023machine,Duda2001}}. Recurrent neural networks enable feedback from the hidden layers such that small transients in the environment that significantly impact the CSI amplitude do not translate into false detections. This is shown at each state by the hidden layer operating as a function of both the previous and current input
\begin{equation}
h_t = \sigma \left( W_{hh}\cdot h_{t-1} + W_{xh}\cdot x_t \right)
\end{equation}
where $x_t$ is the current input, $W_{hh}$ is the current hidden layer neuron, $W_{xh}$ is the input layer neuron, and $\sigma$ is the binary initial activation function. The previous inputs then impact the true output $y_t=W_{hy}\cdot h_t$ weighted by the output layer $W_{hy}$. Various recurrent models are then implemented in a many-to-one configuration, where each input is a measure of the current received CSI magnitude spectrum.

The models are then constructed, with each initial implementation containing a single neural hidden layer and a single dense layer (where every neuron is connected to every neuron of the previous layer) to reduce computational complexity given the strongly isolated feature set. The results of a RNN, long-short term memory (LSTM), and gated recurrent unit (GRU) are utilized for the following CSI target classification~{\cite{zollanvari2023machine,Duda2001}}. Following this, to increase the robustness of the models to transients and changes in the environmental multipath, a bi-directional variant of the LSTM and GRU were introduced, allowing for $y_t$ to be weighted on feature deviations. The models were then trained on a sample batch size of 16 over 25 epochs. The results of these models are shown in Table.~\ref{tab:static_training}, showing high accuracy of identifying the individual transponders, with an accuracy $\geq99\%$ within less than 10 epochs due to the high feature isolation of the five tested cases as shown in Fig.~\ref{fig: accuracies}.

\subsection{Gesture Classification}

While gesture classification using micro-Doppler signatures is a well explored concept \cite{chen2019micro}, using CSI for similar practices in integrated sensing and communications (ISAC) applications is less explored, particularly when multiple devices are present in the field of view.
This work leverages joint classification of the transponder and gesture to reduce ambiguities associated with multiple targets. The classification of gestures with CSI thus requires two key considerations: the stability of spectra classification for dynamic motion and classification of extracted gestures. Classification can be obtained in a training set that combines static and dynamic states. PCA is applied similarly to the previous implementation as shown in Fig.~\ref{fig: pca}(b), where it can be seen that the Mahalanobis distance (MD) increased significantly for all transponders, limiting the maximum possible tag population. However, each tag still largely remains linearly separable, retaining the ability to perform precise identification due to $90\%$ of the spectral content retaining an average MD of $0.001$ obtained from the PCA space of each class. This is further demonstrated when observing the training and validation accuracies of the models under dynamic conditions, shown in Table~\ref{tab:static_training} where a minimum validation accuracy of $80.1\%$ in the single-layer RNN was obtained while multi-layer and bi-directional variants achieved $\geq99\%$ across all classes.  

The motion classification approach leverages traditional micro-Doppler classification by utilizing a convolutional neural network (CNN)~{\cite{11075563,11661345,10965358}}. This approach considers each sample to be the STFT of the first principal component of the calibrated CSI power spectral density over a defined period and Fourier transform length. 
The model resolution of the spectrogram is limited by the observation duration $T$, and thus by the maximum periodicity of the desired classifiable motionm which is proportional to $1/T$. 
The cases evaluated in this work exhibit a maximum periodicity of 0.25 Hz and as such requires an observation period of 4 seconds to sufficiently capture the desired gesture. For CNN classifiers a limiting factor is the available samples and input image resolution~{\cite{zollanvari2023machine,Duda2001}}. However, a large resolution spectrogram is impractical for on-device training in low-cost networks. Thus a series of five convolution and max pooling layers were applied to reduce the input to a $N \times K$ where matrix of $N,K = 256$ for simplified analysis. This reduction is feasible due to the simple features extracted from the periodic motion profiles; more complex gestures may require a longer integration time or increased feature set. The output of the convolution layers was then flattened and passed passed through a dense (fully connected) layer of 128 units, followed by a dropout layer with a rate of 0.5 to prevent overtraining on the limited datasets. The condensed weight matrix was finally passed to a final dense layer corresponding to each of the three gestures. 

\begin{figure}[t!]
\centering
\includegraphics[width = 0.9\linewidth]{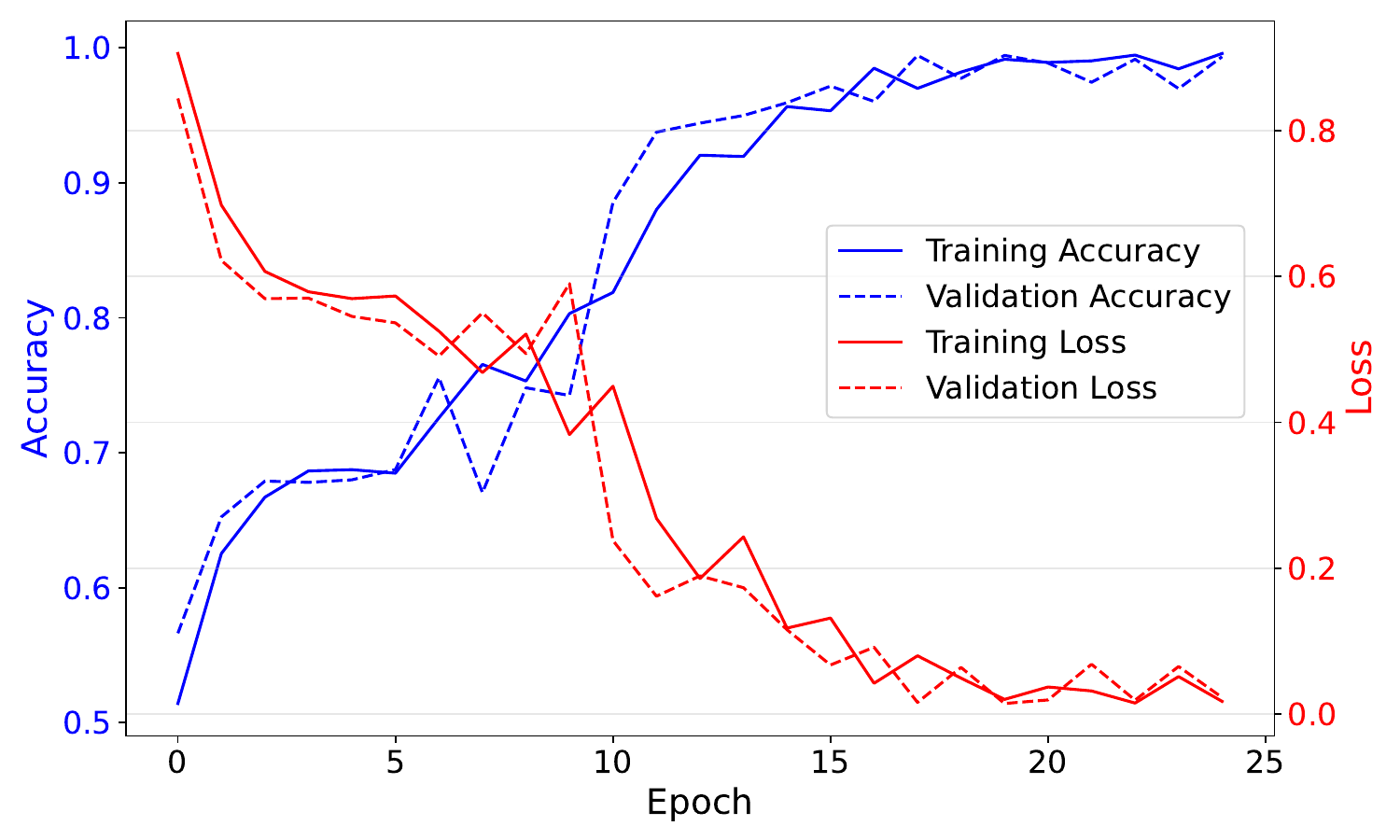}
\caption{Train and test accuracy and loss {for the CNN model with the three known motions as the parameter for supervised learning.}}
\label{fig: CNN}
\end{figure}

The model was trained over a series of 25 epochs to establish the required convergence criteria with a input batch size of 16 samples. As shown in Fig.~\ref{fig: CNN}, the model converges to an accuracy $\geq 99\%$ after only 10 epochs due to the high separation between each of the provided classes. Thus, it can be concluded that with simple deep-learning models that joint gesture classification and device identification can be performed.

\section{Conclusion}
\label{sec: Conclusion}

In this work, we demonstrated that both device identification through nonlinear RF fingerprinting and velocity estimation through CSI measurements can be performed jointly, enabling high-accuracy device identification and gesture classification on commercial hardware. Through a simple analysis of the CSI spectra, it was shown that a unique spectral response can be obtained due to minor differences in the nonlinear transfer function of the components on the transponders. It was also shown that when a nonlinear transponder is in motion the velocity of the desired target can be extracted and classified jointly with the target spectral identifiers. Wjile the experiments in this work focused on the signature of a passive tag equipped with a diode, the approach remains applicable to nonlinear devices in general. 
The simplicity of the tag and system presented highlights the ability to construct efficient, low-cost ISAC systems that leverage existing 802.11 preambles, enabling deployment in existing communications networks with limited computational complexity.

\bibliographystyle{IEEEtran}
\bibliography{Hilton_bib}

\end{document}